\documentclass[aps,twocolumn,english,superscriptaddress,citeautoscript,showkeys,preprintnumbers,amsmath,amssymb,floatfix,footinbib,prb]{revtex4-2}

\usepackage{graphicx}
\usepackage{xcolor}
\usepackage{soul}
\usepackage{amssymb,amsmath}
\usepackage{hyperref}
\hypersetup{colorlinks=true,linkcolor=red,citecolor=blue}
\usepackage[utf8]{inputenc}
\usepackage[english]{babel}
\usepackage{txfonts}
\usepackage{subfig}
\usepackage{siunitx}
\usepackage{booktabs}
\usepackage{placeins}
\usepackage{nicefrac}
\usepackage{caption}
\usepackage{xr-hyper}

\newcommand{\Tc}{$T_\text{c}$}
\newcommand{\dga}{D$\Gamma$A}
\newcommand{\nidxsqysq}{Ni $d_{x^2-y^2}$}
\newcommand{\dxsqysq}{$d_{x^2-y^2}$}

\newcommand*{\addFileDependency}[1]{% argument=file name and extension
\typeout{(#1)}% latexmk will find this if $recorder=0
\@addtofilelist{#1}
\IfFileExists{#1}{}{\typeout{No file #1.}}
}\makeatother

\begin{document}

%\linenumbers

%\title{High-temperature unconventional superconductivity in infinite-layer nickelates at Megabar pressures}

\title{Unconventional superconductivity without doping: infinite-layer nickelates under pressure}

\author{Simone Di Cataldo}\email{simone.dicataldo@uniroma1.it}
\affiliation{Institut f\"{u}r Festk\"{o}rperphysik, Technische Universit\"{a}t Wien, 1040 Wien, Austria} 
\author{Paul Worm} 
\affiliation{Institut f\"{u}r Festk\"{o}rperphysik, Technische Universit\"{a}t Wien, 1040 Wien, Austria} 
\author{Jan M.\ Tomczak} 
\affiliation{King's College London, London, WC2R 2LS, United Kingdom} 
\affiliation{Institut f\"{u}r Festk\"{o}rperphysik, Technische Universit\"{a}t Wien, 1040 Wien, Austria} 
\author{Liang Si} 
%\affiliation{Institut f\"{u}r Festk\"{o}rperphysik, Tecnische Universit\"{a}t Wien, 1040 Wien, Austria} 
\affiliation{School of Physics, Northwest University, Xi'an 710127, China} 
\author{Karsten Held} 
\affiliation{Institut f\"{u}r Festk\"{o}rperphysik, Technische Universit\"{a}t Wien, 1040 Wien, Austria}

\begin{abstract}
High-temperature unconventional superconductivity quite generically 
emerges from doping a strongly correlated parent compound, often (close to) an antiferromagnetic insulator. The recently developed dynamical vertex approximation  is a state-of-the-art technique that has quantitatively predicted the superconducting dome of nickelates. Here, we apply it
to study the effect of pressure in  the infinite-layer nickelate Sr$_x$Pr$_ {1-x}$NiO$_2$. We reproduce the increase of the critical temperature ($T_c$) under pressure found in experiment up to 12 GPa.
According to our results, $T_c$ can be further increased with higher pressures.
%increases further with pressure. 
Even without 
Sr-doping  the parent compound, PrNiO$_2$, will become a high-temperature superconductor thanks to a strongly enhanced  self-doping of the \nidxsqysq{}  orbital under pressure. 
With a maximal \Tc{} of 100\,K around 100\,GPa, nickelate superconductors can reach that of the best cuprates.
\end{abstract}

\date{\today}

\pacs{}

\maketitle
\section{Introduction}
Ever since the discovery of high-temperature superconductivity in LaBaCuO$_{4}$ \cite{Bednorz_JPCMB_1986_BaLaCuO}, understanding or even predicting new  \textit{unconventional} (not electron-phonon-mediated) superconductors and identifying the pairing mechanism has been the object of an immense research effort.
A new opportunity for a more thorough understanding  arose with the discovery of superconductivity in several infinite-layer nickelates $A_{1-x}$$B_{x}$NiO$_2$ \cite{Li_Nature_2019_supernickel, Osada_PRM_2020_PrNiO2_pd, Li_PRL_2020_dome_nickel, Zeng_SciAdv_2021_super_CaNiO2, Osada_AdvMat_2021_LaSrNiO2, Pan_NatMat_2021_5L_nick}, where $A$=La, Nd, Pr and  $B$=Sr, Ca are different combinations of rare-earths and alkaline-earths. These nickelates are at the same time strikingly similar to cuprates (for this reason theory predicted  nickelate superconductivity 20 years before  experiment \cite{Anisimov1999}) but also decidedly different. This constitutes an ideal combination to clarify the presumably common mechanism behind superconductivity in both systems.
Superconductivity in nickelates was found to be quite independent of the rare earth $A$ and dopant $B$ \cite{Chow_FrontPhys_2022_review_nickelates}  with a dome-like shape characteristic of unconventional superconductors.

Theoretical work has left little doubt that nickelates are, indeed, unconventional superconductors \cite{Botana_arXiv_2023_nickel_GWep, DiCataldo_arXiv_2023_LaNiO2_H,Nomura_PRB_2019_NiO2_ep} and the similarity to the crystal and electronic structure of cuprates is striking \cite{Anisimov1999,Held2022}.
 There are, however,  subtle differences between cuprates and nickelates: Compared to Cu$^{2+}$, the  3$d$ bands of Ni$^{1+}$ are separated by a larger energy from the oxygen ones, hence hybridization is weaker and oxygen plays a less prominent role than in cuprates. On the other hand,  the rare-earth $A$-derived bands cross the Fermi level in nickelates and form electron pockets. 
 These electron pockets self-dope the \nidxsqysq{} band with about 5\% holes \cite{Nomura_PRB_2019_NiO2_ep,Botana2019,PhysRevB.102.220501,PhysRevLett.125.077003,Wu2019,jiang2019electronic,Motoaki2019,Kitatani_npj_2020_renaissance}, and prevent the parent compound from being an antiferromagnetic insulator. Given the inherent difficulty to incorporate the effect of strong electronic correlations, different theory groups have arrived at a variety of models for describing nickelates \cite{PhysRevLett.125.077003,Motoaki2019,Kitatani_npj_2020_renaissance,Karp2020,Pascut2023,Adhikary2020,Lechermann2019,Petocchi2020,PhysRevB.108.155147}.
 
% and nickel are such that in the latter a single-band Hubbard model fitted to the \nidxsqysq{} band is sufficient to describe the superconducting properties \cite{Nomura_PRB_2019_NiO2_ep}.

Based on a minimal model consisting of a 1-orbital Hubbard model for the \nidxsqysq{} band \cite{Nomura_PRB_2019_NiO2_ep} plus largely decoupled electron pockets, that only act as electron reservoirs, Kitatani {\em et al.} \cite{Kitatani_npj_2020_renaissance} accurately predicted the superconducting dome in Sr-doped NdNiO$_2$ \cite{Kitatani_npj_2020_renaissance} prior to experiment \cite{Li_PRL_2020_dome_nickel, Li_PRL_2020_dome_nickel, lee2022character}. In  particular, the agreement to more recent defect-free films \cite{lee2022character} is excellent. This includes the quantitative value of \Tc, the doping region of the dome and the skewness of the superconducting dome,
 see Ref.~\cite{Worm2021c}. Also pentalayer nickelates \cite{Pan2021} seamlessly fit the results of Ref. \cite{Worm2021c}.
In Refs.~\cite{Kitatani_npj_2020_renaissance, Kitatani_PRL_2022_palladates} some of us pointed out that  larger  \Tc's should be possible if the ratio of interaction to hopping, $U/t$, is reduced. 
% I think if we mention pressure we have to explain why, or we explain it later
%One way to achieve this is pressure, as it increases the orbital overlap between different atoms and  hence $t$; while $U$   is expected to be rather insensitive to pressure.

In a recent seminal paper, Wang {\em et al.}~\cite{Wang_NatComm_2022_press_PrSrNiO2} reported 
a substantial increase of  \Tc{} in Sr$_{x}$Pr$_{1-x}$NiO$_2$ ($x = 0.18$) films on a SrTiO$_3$ (STO)
substrate from 18\,K to 31\,K if a pressure of\,12 GPa is applied in a diamond anvil cell. There are no indications of a saturation of the increase of \Tc{} with pressure yet.
First calculations of the electronic structure,
fixing the in-plane lattice constant to the ambient pressure value and 
relaxing (reducing) the out-of-plane $c$-axis
%, possibly motivated by one of the explanations for the enhanced \Tc{} given in \cite{Wang_NatComm_2022_press_PrSrNiO2},
have been presented 
\cite{Christiansson_PRB_2023_PrSrNiO2_pressure}. 
A large \Tc{} under pressure \cite{Sun2023} or at least a resistivity drop 
\cite{Sakakibara2023}, has also been reported
in another nickelate: La$_3$Ni$_2$O$_7$. With a 3$d^{7.5}$ electronic configuration and prevalent charge density wave fluctuations the mechanism in this compound is however clearly distinct from the (slightly doped) 3$d^9$ nickelates considered here.

%\ls{I propose a discussion on the three possible ways to enhance Tc in nickelate, as suggested by the conclusions from our NPJ. These include reducing substrate, applying pressure, and replacing 3d with 4d. Among these options, pressure is experimentally more feasible and can be gradually tuned. Therefore, it could shed more light on understanding the pairing force of superconductivity in nickelate when compared to the alternatives of 4d substitution and substrate reduction.}
%\kh{I agree this is interesting, but maybe to close to our npj qu. mat with Motoharu?}
%\ls{Then we can use a different way to say it: Two fundamental questions persist: (1) What is the underlying mechanism driving the pressure-enhanced Tc? and (2) Can an increase in pressure further elevate Tc, potentially surpassing the boiling point of liquid nitrogen?}
% Liang, that's a good point. I put it to the conclusion, now the overall paper is close to the limt already

\begin{figure*}[tb]
\begin{minipage}{.73\textwidth}
\centering
	\includegraphics[width=\columnwidth]{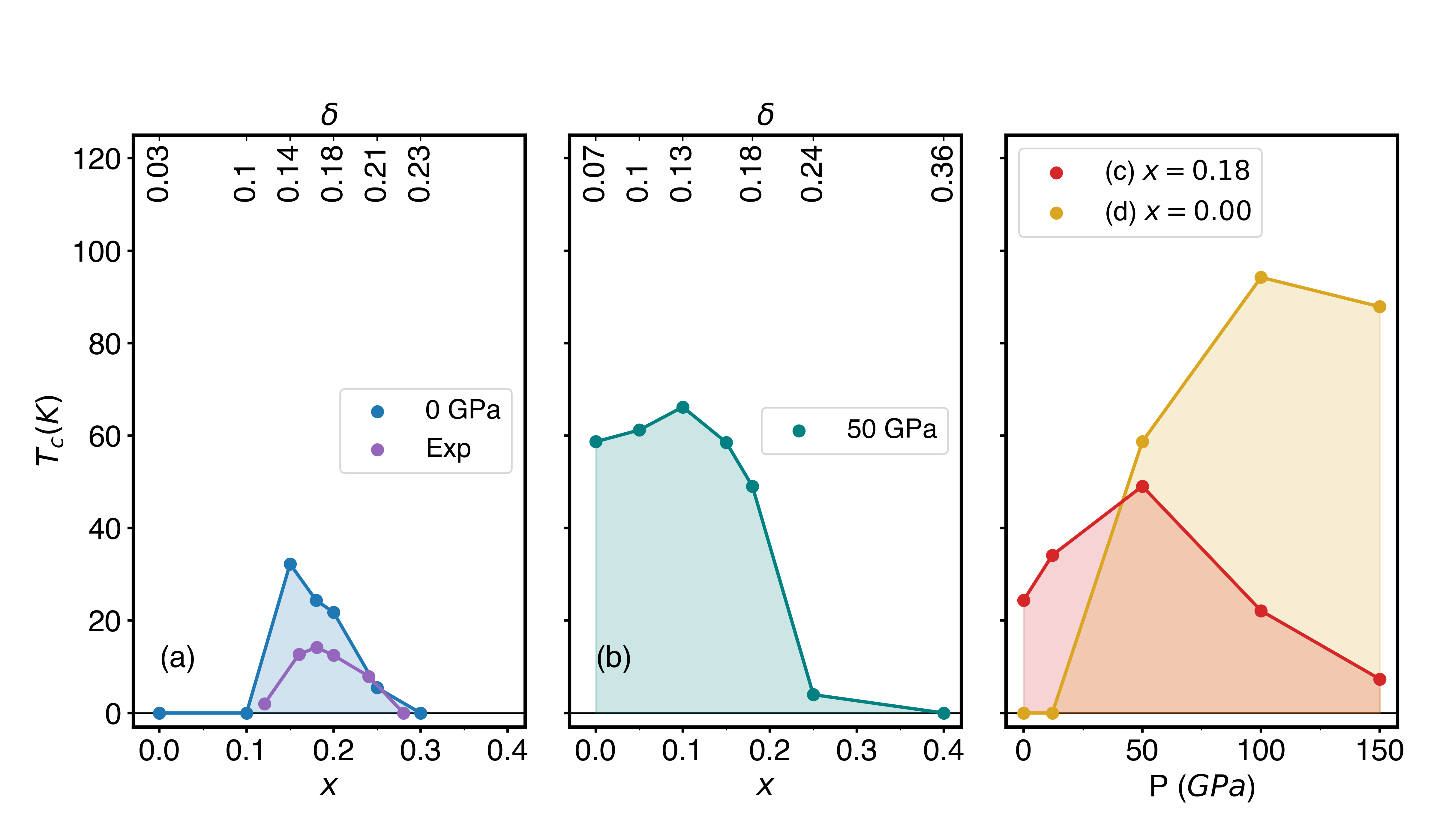}
 \end{minipage}
\hfill
\begin{minipage}{.264\textwidth}
        \caption{Phase diagram  \Tc{}  vs.\ Sr-doping $x$ and  pressure $P$
        of Sr$_x$Pr$_{1-x}$NiO$_2$ as calculated in D$\Gamma$A.  Four different paths are considered: 
          as a function of $x$ for  (a) 0\,GPa and (b) 50\,GPa; as a function of $P$ at  (c) $x = 0.18$ and (d) $x = 0$. The secondary (upper) $x$-axis of panels (a), (b) shows the effective hole doping $\delta$ of the \nidxsqysq{} orbital with respect to half filling. Panel (a) also compares to the experimental result from Ref. \cite{Osada_NanoLett_2020_PrNiO2} (purple dots). }
 	\label{fig:tc_summary}
  %\kh{same y-axis for (a) and (b) would be nice and some shading to make the phase diagrams  look more fancy}
  \end{minipage}
\end{figure*}
In this work, we employ  the same state-of-the-art scheme that was so successful in Ref.~\cite{Kitatani_npj_2020_renaissance}, which is based on density functional theory (DFT), dynamical mean-field theory (DMFT), and the dynamical vertex approximation (D$\Gamma$A). We study the pressure dependence of the superconducting phase diagram of  PrNiO$_2$ (PNO) with and without Sr doping. 
We find (i) a strong increase of  the hopping $t$, almost by a factor of two, when going from 0 to 150\,GPa, while the value of $U$, obtained through constrained random-phase approximation (cRPA), remains essentially unchanged as in cuprates \cite{Upressure}. 
Importantly, pressure further results in (ii) deeper electron pockets, effectively increasing the hole doping $\delta$ of the \nidxsqysq{} band with respect to half-filling. 
Altogether, this results in the phase diagrams shown in Fig.~\ref{fig:tc_summary}, the main result of our work.
When going from (a) ambient pressure to (b) 50\,GPa, \Tc{} increases by up to a factor of two and $d$-wave
superconductivity is observed in a much wider doping range -- quite remarkably even without Sr doping.
As a function of pressure, at doping fixed to $x=0.18$ (c), the simulated phase diagram
%The phase diagram 
%as a function of pressure shows %, at $x=0.18$ (c), 
shows a very similar increase of \Tc{}  from 0 to 12\,GPa as in experiment \cite{Wang_NatComm_2022_press_PrSrNiO2}.  
The figure further reveals that 
\Tc{} will continue to increase up to 49 K at 50\,GPa, followed by a rapid decrease at higher pressures. 
For the parent compound, PrNiO$_2$ ($x=0$, d), 
the enhanced self-doping alone is sufficient to turn it superconducting  with a maximum predicted \Tc{} of close to 100$\,$K at 100\,GPa.

\section{Results}
\label{sect:results}
As the superconducting nickelate films are grown onto a STO substrate, 
particular care has to be taken when simulating the effect of 
%we model the effect of 
isotropic pressure in the diamond anvil cell
%as closely as possible to the real system
(cf.\ Supplemental Material (SM) \cite{SM} Section~\ref{supp:subsect:methods_workflow} for a flowchart of the overall calculations and Section~\ref{supp:subsect:methods_relaxation} for further details of the pressure calculation).
First, since the  thickness of the film is  10-100\,nm  and thus negligible compared to that of the STO substrate, we calculate 
the STO equation of state in DFT and, from this, obtain the STO lattice parameters under pressure. 
Second, we fix the in-plane $a$ (and $b$) lattice parameters to that of STO 
under pressure
and find the lattice parameter $c$ for the nickelate which minimizes the enthalpy at the given pressure. The resulting lattice constants are shown in Table~\ref{tab:hoppings}.
This procedure better reflects the response of the system to the rather isotropic pressures realized in experiment and is more realistic than that used
%notably different from that
in Ref.~\cite{Christiansson_PRB_2023_PrSrNiO2_pressure}
where the $a$-$b$ lattice parameters had been fixed 
to that of unpressured STO \cite{Christiansson_PRB_2023_PrSrNiO2_pressure}.

\begin{table}
    \centering
    \begin{tabular}{c|c|c|c|c|c}
        $P$ & $a$  & $c$  & $t$ & $t'$ & $t''$   \\
        
        [GPa] & [\AA] & [\AA] & [eV] & [eV] & [eV]  \\
        \hline
        0    & 3.90  & 3.32 & -0.39  & 0.10 & -0.05  \\
        12.1 & 3.83 & 3.20 & -0.42 & 0.10  & -0.05   \\
        50   & 3.67  & 3.03  & -0.48 & 0.11 & -0.06 \\
        100  & 3.54  & 2.89  & -0.56 & 0.11 & -0.07\\
        150  & 3.45  & 2.79  & -0.62 & 0.12  & -0.07 
    \end{tabular}
    \caption{{\it Ab initio} values for the lattice constants and the hoppings of the 1-orbital Hubbard model for PrNiO$_2$ under pressure. Here,
    $t$, $t'$, and $t''$ are the nearest, next-nearest, and next-next-nearest neighbour \dxsqysq-hoppings.}
    %All energies are in units of eV.}
    \label{tab:hoppings}
\end{table}

\begin{figure*}[htb]
\centering
	\includegraphics[width=1.8\columnwidth]{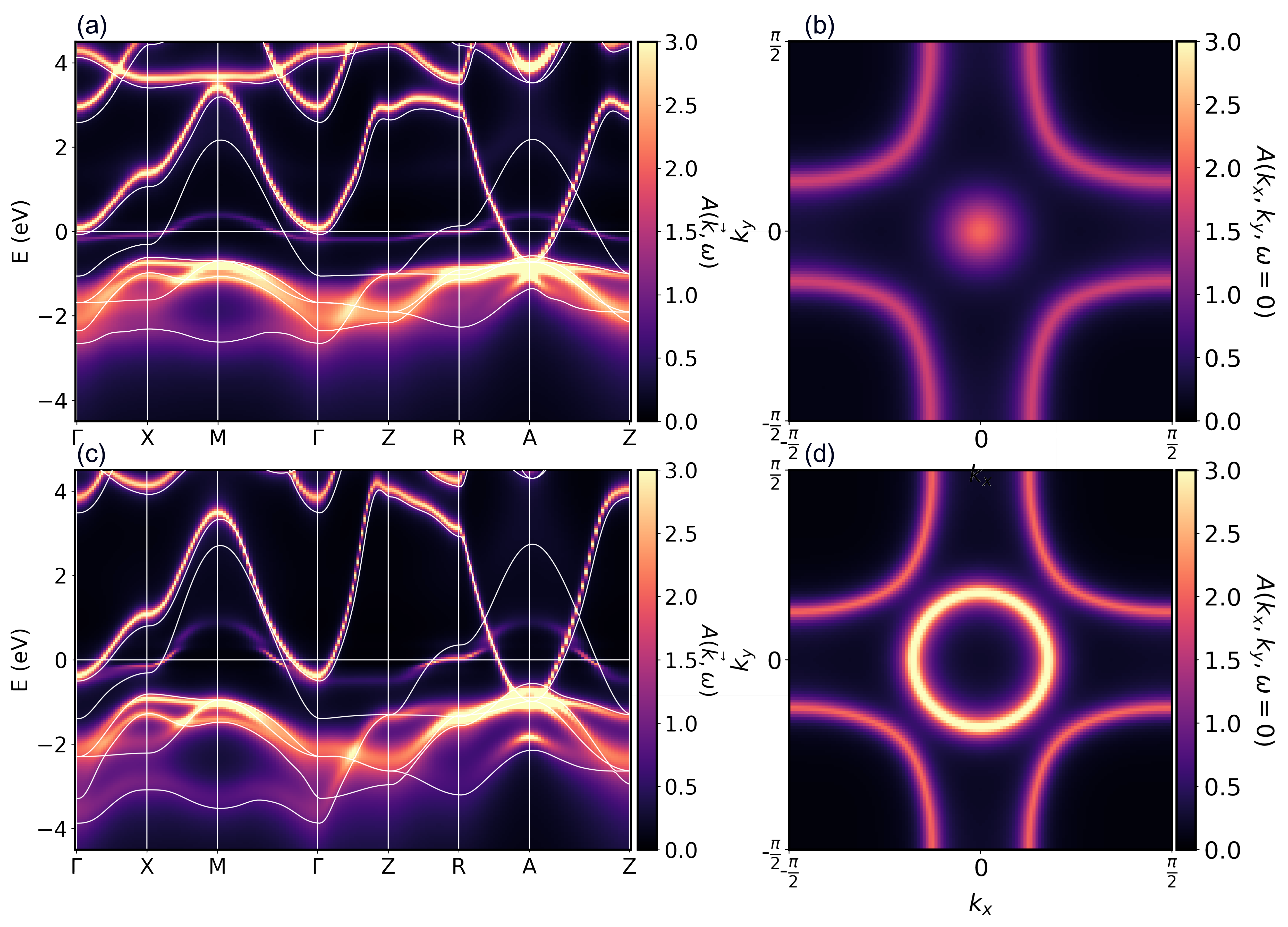}
        \caption{Spectral function for undoped PNO. Panels (a) and (c) show the DMFT spectral function $A(\textbf{k}, \omega)$ (color scale) and the Wannier bands (white lines) along a path through the Brillouin zone  at temperature $T = 300 K$. Panels (b) and (d) show the same spectral function in the $k_z = 0$ plane. $A(k_x, k_y, k_z=0, \omega = 0)$. The $\Gamma$ point is at the center, i.e.~$(k_x=0,k_y=0)$.
     %O.K.
        %\sdc{For 10 bands calculation I used physical units, including a fixed temperature. The 1 band calculations I used units of t and fixed beta. Conversion is 300 K -> beta = 1/T = 15-18 in units of t} Panels (b) and (d) show the spectral function of the Fermi surface (energy zero) in the $k_{z} = 0$ plane. \kh{somehow the caption is centered not left side aligned}}
        }
 	\label{fig:1}
\end{figure*}

With the crystal structure determined, we 
calculate the DFT electronic structure at pressures of 0, 12, 50, 100, and 150\,GPa.  Next, we perform a 10- and 1-orbital Wannierization around the Fermi energy, including all Pr-$d$ plus Ni-$d$ orbitals and only the \nidxsqysq{} orbital, respectively. The DFT band structures and Wannier bands 
are shown in SM ~\cite{SM} Fig. \ref{supp:fig:wannier_pressure} and as white lines in Fig.~\ref{fig:1}.

Following the method of Refs.~\cite{Kitatani_npj_2020_renaissance,Kitatani2022}, we supplement the Wannier Hamiltonian with a local intra-orbital Coulomb interaction of $U=4.4\,$eV (2.5\,eV) and Hund's exchange $J=0.65\,$eV (0.25\,eV) for Ni-3$d$ (Pr-5$d$) as calculated in cRPA~\cite{Si_PRL_2020_LaNiO2_H}.
For the thus derived 10-band model we
perform DMFT calculations.
The resulting DMFT spectral function of undoped PNO 
is shown in Fig.~\ref{fig:1}, for 0 and 50\,GPa. We see that the \nidxsqysq{} orbital crossing the Fermi energy is strongly quasiparticle-renormalized compared to the DFT result (white lines).
In addition, there are pockets at the $\Gamma$ and A momenta, which essentially follow the DFT band structure without renormalization.

Important for the following is that, with the overall increase of bandwidth under pressure, the size of the pockets  grows dramatically under pressure in DFT and DMFT alike.
The enlargement of the $\Gamma$ pocket can also be seen from the Fermi surface Fig.~\ref{fig:1} (d) vs.\ (b). The effect at higher pressures is shown in  SM ~\cite{SM}  Fig. \ref{supp:fig:10_bands_0_doping_vs_pressure}-\ref{supp:fig:fs_pressure}.

The results of the 10-band model show that the low-energy physics of the system boils down to one-strongly correlated \nidxsqysq{} orbital plus weakly correlated electron pockets. Here, the A-pocket does not hybridize by symmetry with the \nidxsqysq{} band,
as is evident by the mere crossing of both between R and A. The band forming the
$\Gamma$ pocket, on the other hand, mainly hybridizes with other Ni orbitals that in turn couple to the \nidxsqysq{} through the Hund's $J$.  Note, however, that the main spectral contribution of these other Ni-$d$ orbitals is still well below the Fermi energy in Fig.~\ref{fig:1}.

The above justifies a one-band minimal model for superconductivity in PNO
\cite{Kitatani_npj_2020_renaissance,Kitatani_PRL_2022_palladates,Held2022}, with the working hypothesis that superconductivity arises from the correlated \nidxsqysq{} band only. 
However, the effective hole-doping $\delta$ of the \nidxsqysq{} band (relative to half-filling) has to be calculated from the 10-band DFT+DMFT to properly account for the electrons in the pockets. In the following, it is thus imperative to always distinguish between the number of holes corresponding to Sr substitution of the Pr site (\textit{chemical} doping $x$) and the holes in the \nidxsqysq{} band compared to half filling (\textit{effective} doping $\delta$).

The electron pockets induce a nonlinear dependency of $\delta$ from $x$, and their growth with pressure $P$ causes $\delta$ to increase by about 0.06 from 0 to 100\,GPa, see SM~\cite{SM} Fig.~\ref{supp:fig:pressure_eff_filling}.
 
 Using the effective doping $\delta$ of the \nidxsqysq{} band, we perform a second DMFT calculation for the single \nidxsqysq{} orbital, which we describe as a single-band Hubbard model with an interaction of $U=3.4\,$eV. This $U$ is smaller than for the 10-band model due to additional screening, but it is notably insensitive to pressure (cf.\ SM~\cite{SM} Tab.~\ref{supp:tab:U}). The main effect of pressure is instead the increase of $t$ as summarized in
 Table \ref{tab:hoppings} and the already mentioned enhanced self-doping. 
 %The interested reader can find further details, including the change of the hopping parameters, in SM~\cite{SM} Section~\ref{...} and in Table \ref{tab:hoppings}.

\begin{figure}[tb]
\centering
 \vspace{-1.5cm}
 
	\includegraphics[width=0.95\columnwidth]{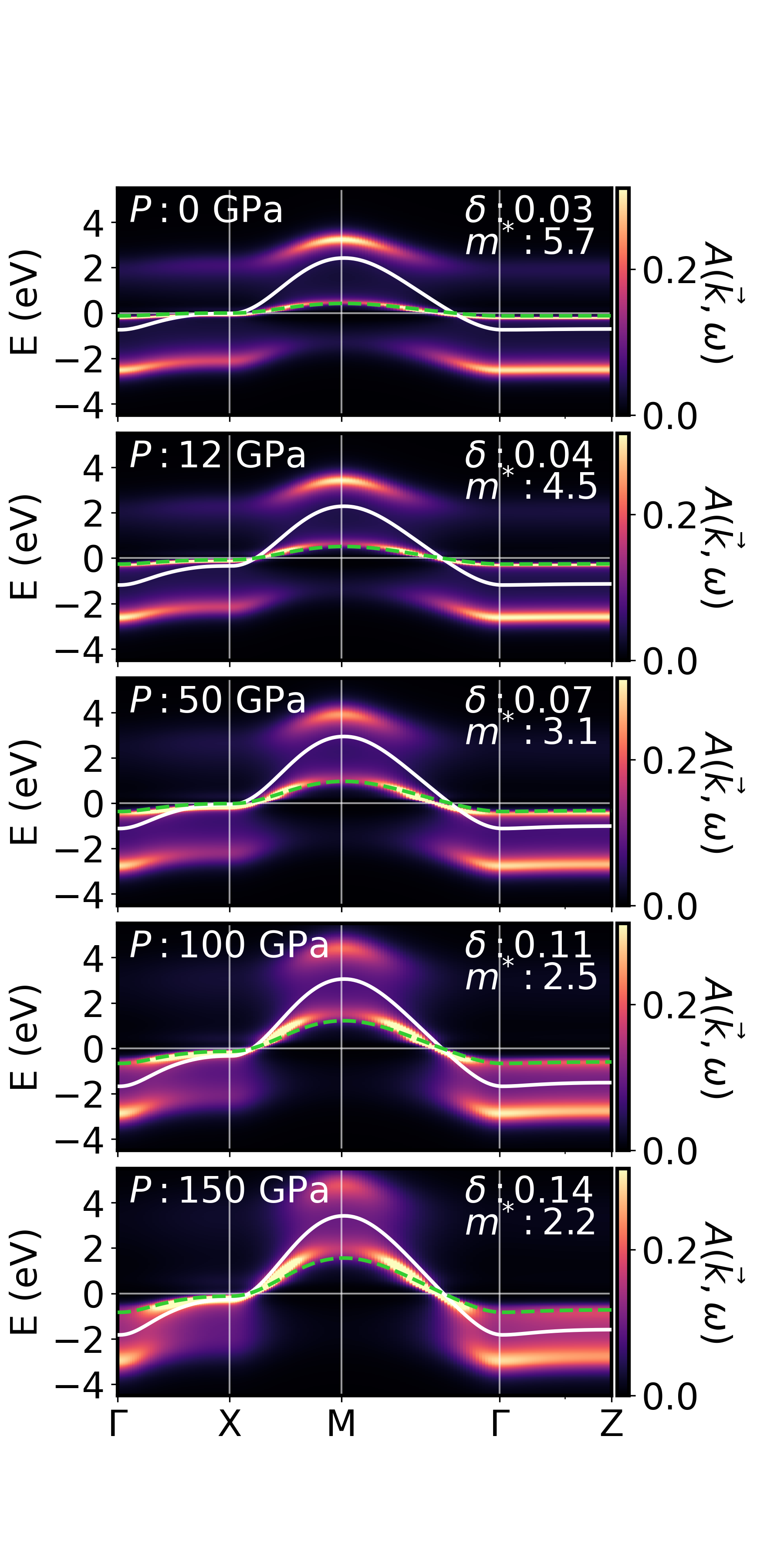}
 \vspace{-1cm}
        \caption{DMFT spectral function (color bar) for undoped PNO for the 1-band model and as a function of pressure from 0 to 150\,GPa at $T=0.0125 t$. The original DFT Wannier band and the same band renormalized with the effective mass are shown as solid white and dashed green lines, respectively. }
 	\label{fig:spfun_local_pressure}
\end{figure}

  In Fig. \ref{fig:spfun_local_pressure}, we show the spectral function $A(\textbf{k}, \omega)$ of the 1-band model for PNO as a function of pressure and $x = 0$.
  The panels for 0\, and 50\,GPa can be compared to Fig.~\ref{fig:1} (a,c) and show that the 1-band model reproduces the renormalization of the
  \nidxsqysq{} orbital in the fully-fledged 10-band calculation.

  Hubbard bands are visible at all pressures in Fig. \ref{fig:spfun_local_pressure}, but become more spread out and less defined with increasing pressure. Simultaneously, the effective mass $m^{*}$ decreases, and the bandwidth widens.  Similar results but for the experimentally investigated Sr-doping 
  $x = 0.18$ can be found in SM~\cite{SM} Fig.~\ref{supp:fig:spfun_d_0.18_pressure},
  and as a function of doping at 50\,GPa in Fig.~\ref{supp:fig:spfun_doping_50_GPa}.

   Next, we calculate the  superconducting \Tc{} using D$\Gamma$A~\cite{Rohringer2018,Kitatani2022}. 
     In Fig.~\ref{fig:tc_summary}, we follow four different paths in parameter space: as a function of doping at (a) 0\,GPa and (b) 50\,GPa as well as  as a function of pressure with Sr-doping  (c) $x=0.18$, i.e., for the parent compound, and (d)  $x=0$.
     At ambient pressure in Fig.~\ref{fig:tc_summary}~(a) our results are in excellent agreement with our previous calculations for other nickelates \cite{Kitatani_npj_2020_renaissance, Kitatani_PRL_2022_palladates}, cf.~SM~\cite{SM} Fig.~\ref{supp:fig:tc_and_t_p}. The small differences are ascribable to the slightly different material, and are in good agreement with experiment \cite{Osada_PRM_2020_PrNiO2_pd,lee2022character}. We find the effect of pressure to be significant: At 50\,GPa, the maximum \Tc{} is enhanced by a factor of two compared to ambient pressure, while the maximum slightly shifts to lower doping ($x = 0.10$). 
     Remarkably, even the undoped parent compound  becomes superconducting at 50~GPa [$x=0$ in Fig.~\ref{fig:tc_summary} (b)], due to the increased self-doping from the electron pockets.

    The experimental Sr-doping of  $x$=0.18 \cite{wang2021pressure}  is close to optimal doping at 0\,GPa. Increasing pressure in Fig.~\ref{fig:tc_summary}~(c),
    we observe an increase of   
    \Tc{}  by 0.81 K/GPa in excellent agreement with the experimental  rate of 0.96 K/GPa \cite{wang2021pressure}, for pressures  up to 12\,GPa. 
    The predicted \Tc{} of 30\,K at 0\,GPa is slightly higher in theory than the experimental 18\,K \cite{wang2021pressure}, but still in good agreement.  As pressure increases beyond 12\,GPa, \Tc{} continues to grow and peaks at around 50 GPa with 49\,K, before decreasing for higher pressures.
    
    Most striking is the result for the undoped compound PNO in Fig.~\ref{fig:tc_summary}~(d). Here, superconductivity sets in below 50\,GPa and peaks at almost 100\,K around 100\,GPa.  Intrinsic doping from the electron pockets is sufficient to make the parent compound superconducting at high temperature.
    
\section{Discussion}
\label{sect:discussion}
\begin{figure}[tb]
\centering
	\includegraphics[width=0.95\columnwidth]{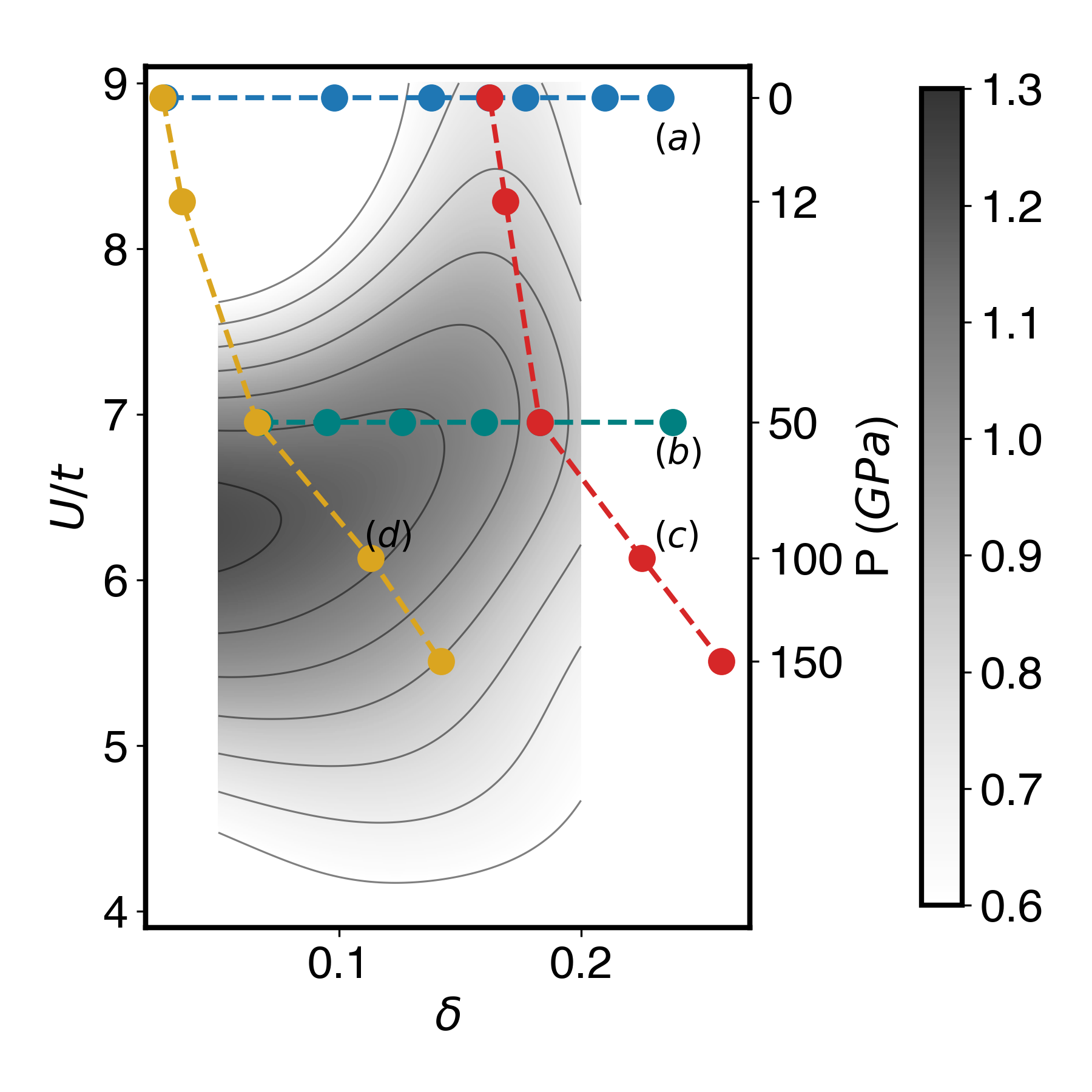}
        \caption{
        The considered four paths  in the $U/t$ 
        vs.\ effective  hole doping $\delta$ 
        parameter space:
        at a fixed pressure of (a) 0\,GPa and (b) 50\,GPa
        as a function of Sr-doping $x$;
        as a function of pressure for fixed Sr-doping (c) $x=0.18$ and (d) $x=0.00$. The gray color bar
    indicates the strength of superconductivity (superconducting eigenvalue $\lambda$  at $T = 0.01$t; from~\cite{Kitatani_PRL_2022_palladates}). The secondary $y$ axis reports the pressure corresponding to the $U/t$ values shown.}
 	\label{fig:summary_figure}
\end{figure}

To rationalize our results, we plot 
the four paths at fixed pressure, respectively, fixed Sr-doping $x$ in Fig.~\ref{fig:summary_figure}, but now 
as a function of $U/t$
and the effective hole doping $\delta$
of the \nidxsqysq{} orbital. 
Superimposed is the D$\Gamma$A superconducting eigenvalue $\lambda$ \cite{Kitatani_PRL_2022_palladates}, with the darker gray regions corresponding to a higher \Tc.

The application of an isotropic pressure on infinite-layer PrNiO$_2$ has two effects: First, it boosts the hopping $t$ of the \nidxsqysq{} orbital, which at 150\,GPa becomes almost twice as large than at 0\,GPa, see Table \ref{tab:hoppings}.
This increases the overall energy scale and thus enhances \Tc. Since $U$ does not change significantly,
the ratio $U/t$ also decreases.
This is preferable for superconductivity since at ambient pressure PNO exhibits a $U/t$ slightly above the optimum of $U/t \sim 6$ (above the darker gray region in Fig.~\ref{fig:summary_figure}). However, at high pressures of e.g.\ 100\,GPa and 150\,GPa curves (c) and (d) have passed the optimum in Fig.~\ref{fig:summary_figure}; \Tc{} in Fig.~\ref{fig:tc_summary} decreases again.

Second, pressure enhances the effective hole doping $\delta$ even at fixed Sr-doping $x$, as the electron pockets become larger. For this reason, curves (c) and (d) in Fig.~\ref{fig:summary_figure} deviate from a vertical line. For Sr-doping (c) $x=0.18$, which
is close to optimum at 0\,GPa, the curve moves away from optimum doping to the overdoped region when pressure is applied. This is a major driver for the decrease of \Tc{}  above 50\,GPa in  Fig.~\ref{fig:tc_summary} (c).
In stark contrast, for the parent compound ($x=0$; d) the effective hole doping $\delta$ goes from underdoping to optimal doping, and only at much larger pressures to overdoping and too small $U/t$.
Consequently, PNO without doping hits the sweet spot 
for superconductivity in Fig.~\ref{fig:summary_figure}
at a pressure between 50 and 100\,GPa.

\section{Conclusion}
In short, our results strongly suggest that experiments for infinite-layer nickelates are still far from having achieved their maximum \Tc{}. Surprisingly, the maximum \Tc{} of almost 100\,K 
is predicted to be found in  undoped PrNiO$_2$ between 50 and 100\,GPa. This 
places nickelates almost on par with cuprates in the Olympus of high-\Tc{} superconductors. The nickelate phase diagram under pressure will not only exhibit a significant increase in \Tc{} but also a wider dome. In particular, the maximum of this  dome is shifted to lower Sr-doping $x$ when a pressure of 50 - 100\,GPa is applied. 

Such pressures can be achieved experimentally in diamond anvil cells.
%and, given the higher \Tc, cryogenic cooling with liquid nitrogen is possible.
An alternative route to obtain the same in-plane 
lattice compression is using a substrate with smaller lattice parameters. For example, LaAlO$_3$, YAlO$_3$ and LuAlO$_3$
have lattice parameters 
of 3.788\,\AA~\cite{savchenko1986phase}, 3.722\,\AA~\cite{ismailzade1971x} and 3.690\,\AA~\cite{shishido1995flux}, respectively, which are already close to the in-plane lattice constants at 50\,GPa. As this approach would not change the out-of-plane lattice parameter, the self-doping of the \nidxsqysq{} band from the electron pockets should be less important. Hence we expect that with these substrates a higher \Tc{} might be achieved, but only with at least 10\% doping.
% \sdc{} 

\section{Methods}
In this section, we summarize the computational methods employed.  The interested reader can find additional information in the Supplementary Material \cite{SM}, and data and input files for the whole set of calculations in the associated data repository \cite{suppdata}.

Density functional theory calculations were performed using the Vienna ab-initio simulation package (VASP) \cite{Kresse_PRB_1996_VASP, Kresse_PRB_1999_VASP_pseudos} using projector-augmented wave pseudopotentials and Perdew-Burke-Ernzerhof exchange correlation functional adapted for solids (PBESol) \cite{Perdew_PRL_1996_PBE, PhysRevLett.100.136406}, with a cutoff of 500 eV for the plane wave expansion. Integration over the Brillouin zone was performed over a grid with a uniform spacing of 0.25\,$\AA^{-1}$ and a Gaussian smearing of 0.05\,eV. Wannierization was performed using \verb|wannier90| \cite{mostofi2008wannier90}. 

The Hubbard $U$ of the 1-orbital setup under pressure was computed from first principles using the constrained random phase approximation (cRPA) in the Wannier basis \cite{miyake:085122} for entangled band-structures \cite{miyake:155134}, relying on a DFT electronic structure obtained from a full-potential linearized muffin-tin orbital method \cite{fplmto}.

DMFT calculations were performed using \verb|w2dynamics| \cite{Wallerberger_CompPhysComm_2019_w2dynamics}, with values of $U$, $J$, and $t$ as detailed in the main text, and in Tab. \ref{supp:tab:summary_dmft_inputs} and SM \cite{SM} Tab.~\ref{supp:tab:10band_DMFT}. The 10-band calculations were performed at a temperature of 300\,K, with a total of 30 iterations to converge the local Green's function, and a final step with higher sampling. The 1-band DMFT calculations were performed at variable temperature, with a total of 70 iterations, and a final step with higher sampling, which was increased at lower temperatures. 

The calculation of the non-local quantities via ladder \dga{} and the solution of the linearized Eliashberg equation was performed starting from the local vertex calculated with \verb|w2dynamics| using our own implementation, available upon reasonable request.

\section*{Author Contributions}
S. D. C. performed the DFT, and DMFT calculations. S. D. C. and P. W. performed the \dga{} calculations. J. M. T. performed the cRPA calculations. L. S. and K. H. designed and supervised the project. 
All authors participated in the discussion, contributed to the writing of the manuscript, and approved the submitted version.
\section*{Competing interests}
The authors declare that they have no competing interests.

\section*{Data and materials availability}
The raw data for the figures reported, along with input and output files is available at \cite{suppdata}. Related data including the local Green's functions and DFT calculations are also available in the \verb|NOMAD| repository \cite{nomad_repo, Scheidgen_JOpenSourceSoft_2023_nomad}.
\section*{Acknowledgements}
We would like to  thank Motoharu Kitatani and Juraj Krsnik for helpful discussion. We further acknowledge funding through the Austrian Science Funds (FWF) projects ID I 5398,  I 5868, P 36213, SFB Q-M\&S (FWF project ID F86), and Research Unit QUAST by the Deutsche Forschungsgemeinschaft (DFG; project ID FOR5249). L.~S. is thankful for the starting funds from
Northwest University. Calculations have been done on the Vienna Scientific Cluster (VSC).
For the purpose of open access, the authors have applied a CC BY public copyright licence to any Author Accepted Manuscript version arising from this submission.

%\bibliographystyle{apsrev4-2}
%\bibliography{library,add}
%apsrev4-2.bst 2019-01-14 (MD) hand-edited version of apsrev4-1.bst
%Control: key (0)
%Control: author (72) initials jnrlst
%Control: editor formatted (1) identically to author
%Control: production of article title (-1) disabled
%Control: page (0) single
%Control: year (1) truncated
%Control: production of eprint (0) enabled
%

\newpage
%%%%%%%%%% Merge with supplemental materials %%%%%%%%%%
\pagebreak
\widetext
\begin{center}
	\textbf{\large Supplementary Information: \\ Unconventional superconductivity without doping: infinite-layer nickelates under pressure}
\end{center}
%%%%%%%%%% Merge with supplemental materials %%%%%%%%%%
%%%%%%%%%% Prefix a "S" to all equations, figures, tables and reset the counter %%%%%%%%%%
\setcounter{equation}{0}
\setcounter{figure}{0}
\setcounter{table}{0}
\setcounter{page}{1}
\makeatletter
\renewcommand{\theequation}{S\arabic{equation}}
\renewcommand{\thefigure}{S\arabic{figure}}
\renewcommand{\bibnumfmt}[1]{[S#1]}
\renewcommand{\citenumfont}[1]{S#1}
%%%%%%%%%% Prefix a "S" to all equations, figures, tables and reset the counter %%%%%%%%%%

\maketitle

	This Supplemental information contains additional information on the methods employed in Section~\ref{SSec:methods} and additional results in Sections \ref{SSec:results}.
	%, and further computational details in ~\ref{SSect:computational_details}.
	Specifically, Section~\ref{subsect:methods_workflow}
	provides an overview; Section~\ref{subsect:methods_dft} gives details of the density functional theory (DFT) calculation; Section ~\ref{subsect:struct_relax} gives details of the structural relaxation; Section ~\ref{subsect:wannier_results} gives details on the Wannierization of the DFT bands at the different pressures. Section ~\ref{subsect:methods_hubbard} provides information on the constrained random phase approximation (cRPA) calculation of the interaction parameters; Section ~\ref{subsect:DMFT} gives further details on the dynamical mean-field theory (DMFT) calculations; and Section~\ref{subsect:DGA} on the dynamical vertex approximation (\dga). Additional results are presented in Section~\ref{SSec:results}. This includes DFT+DMFT calculations for the 10- and 1-band model in Section~\ref{subsec:results:DMFT}, a comparison of the \dga{} \Tc{} to that of Sr$_x$Nd$_{1-x}$NiO$_2$ in 
	Section~\ref{subsec:results:DGA}, and an in-depth test of the virtual crystal approximation (VCA) in Section~\ref{subsec:VCAtest}.

\date{\today}

\maketitle

\section{Methodological details}
\label{SSec:methods}
In this section, we provide a more detailed description of the methods employed in the main text.
The interested reader can find the input files for the whole set of calculations in the associated data repository \cite{suppdata}.

\subsection{Complete workflow}
\label{subsect:methods_workflow}
\begin{figure*}[tb]
	\centering
	\includegraphics[width=0.95\columnwidth]{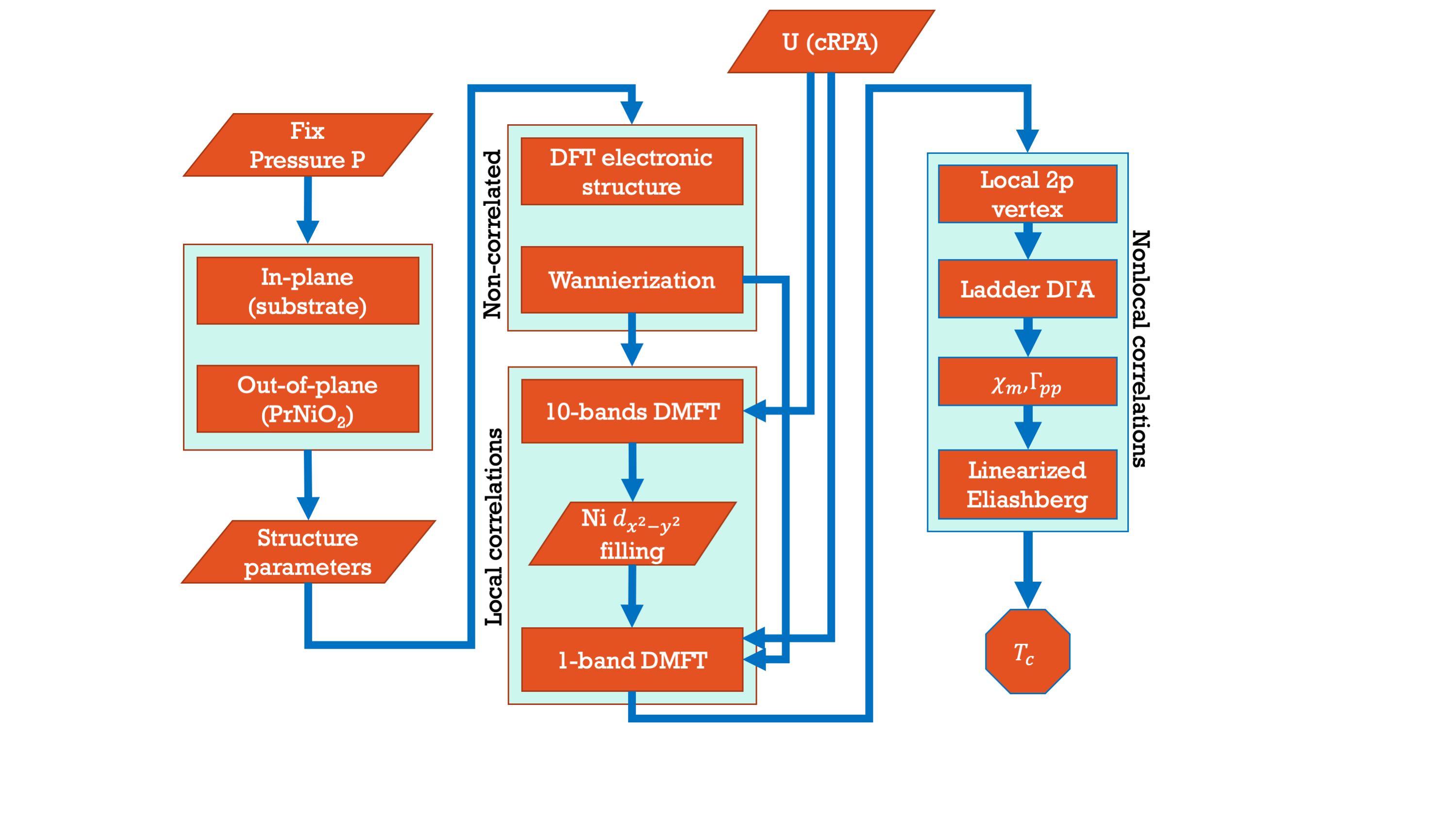}
	\caption{Flowchart summarizing the whole workflow employed in the project.}
	\label{fig:flowchart}
\end{figure*}

We start with presenting a flowchart of the entire methodology as an overview in Fig.~\ref{fig:flowchart}; and further details on the individual steps can then be found in the subsequent Sections.
We start by computing the in-plane lattice parameter as a function of pressure from the STO equation of state in DFT and fix the PNO in-plane lattice parameters to that of the substrate. Next, we
determine the out-of-plane lattice parameter of PNO, as detailed in \ref{subsect:methods_relaxation}. From the relaxed crystal structure, we compute the electronic dispersion $\epsilon_{\textbf{k}}$ using Density Functional Theory and extract the Wannier functions for 10 bands (Ni-$d$ + Pr-$d$) and 1 band (\nidxsqysq{}). 

As described in Ref. \cite{Kitatani_npj_2020_renaissance}, we first perform a DMFT calculation for the 10 bands case, which allows us to obtain the filling of the \nidxsqysq{} in the case of interacting Ni and Nd orbitals. The values of interaction employed in this calculation are reported in  Table \ref{tab:10band_DMFT}.
Using the thus computed filling of the \nidxsqysq{} band, we perform a second DMFT calculation for the single \nidxsqysq{} band, from which we obtain the local two-particle Green's function $G^{(2)}(i\nu, i\nu', i\omega)$.

Finally, using ladder-\dga{} we compute the non-local vertex and the (magnetic) susceptibility at various temperature. The superconducting \Tc{} is then estimated as the temperature for which the leading eigenvalue of the linearized Eliashberg equation is equal to one, as in Ref. \cite{Kitatani_npj_2020_renaissance}.

\subsection{Density Functional Theory}
\label{subsect:methods_dft}
DFT calculations are performed with the Vienna ab-initio Simulation Package (VASP) \cite{Kresse_PRB_1996_VASP}, using the Perdew-Burke-Ernzerhof adapted for solids (PBESol) \cite{PhysRevLett.100.136406} exchange-correlation functional. We employ the projector-augmented waves (PAW) pseudopotentials provided within the VASP package \cite{Kresse_PRB_1999_VASP_pseudos}, for which praseodymium is constructed with $f$ orbitals frozen within the core states. Integration over the Brillouin zone was performed over a uniformly-spaced grid with a spacing of 0.25 \AA$^{-1}$, and a Gaussian smearing of 0.05\,eV. The input and output files for all calculations performed are available as additional data \cite{suppdata}.

At the DFT level, the Sr-doping of the PrNiO$_2$ crystal was simulated by means of the virtual crystal approximation (VCA) to simulate a virtual Pr$_{1-x}$Sr$_{x}$ atom following the method describe by Bellaiche and Vanderbilt \cite{Vanderbilt_PRB_2000_VCA}, and implemented in VASP. The use of this approximation, despite the different core of Sr and Pr is justified by the fact that in the system studied both, Pr and Sr, act mostly as charge donors and spacers between NiO$_2$ planes, and do not contribute significantly to the states at the Fermi energy. Nevertheless, we checked extensively the quality of the VCA against Vergard's law, and on the structural and electronic properties of Pr$_{0.75}$Sr$_{0.25}$NiO$_2$, compared with the results on a 2$\times$2$\times$2 supercell. 
As these checks are quite extensive and we do not want to interrupt the further discussion of the workflow here, we present  in Section~\ref{subsec:VCAtest}, Figures \ref{fig:vergards} (Vergard's law) and \ref{fig:prsrnio2_bands_vca} (comparison with  2$\times$2$\times$2 supercell).
This aspect of  Figure \ref{fig:prsrnio2_vca_test}, which we present already in Section~\ref{subsect:methods_relaxation}  for its information on how to obtain the $c$-axis parameter,  is also discussed in  Section~\ref{subsec:VCAtest}.
All these tests confirmed that the VCA is consistent with the results obtained for supercells in the structure studied.

\subsection{Structural Relaxation}
\label{subsect:methods_relaxation}
\begin{figure}[htb]
	\centering
	\includegraphics[width=0.45\columnwidth]{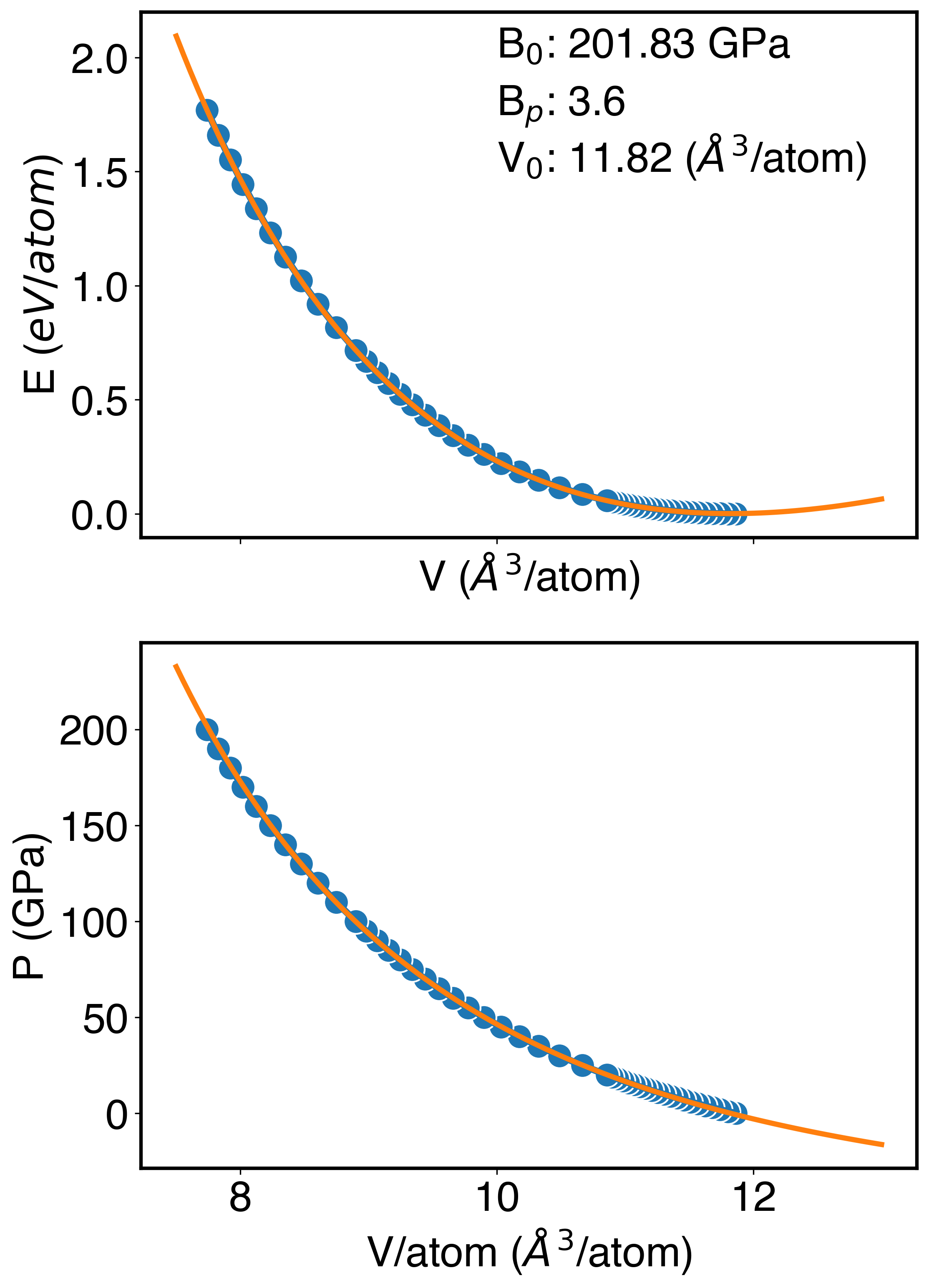}
	\caption{Equations of state $E(V)$ and $P(V)$ for SrTiO$_3$ from 0 to 200 GPa. The blue dots and the orange line denote the values computed within Density Functional Theory and the fit, using the Birch-Murnaghan equation of state \cite{Murnaghan1944}. In addition, we report the results of the curve fit. }
	\label{fig:STO_eos}
\end{figure}

\begin{figure}[htb]
	\centering
	\includegraphics[width=0.45\columnwidth]{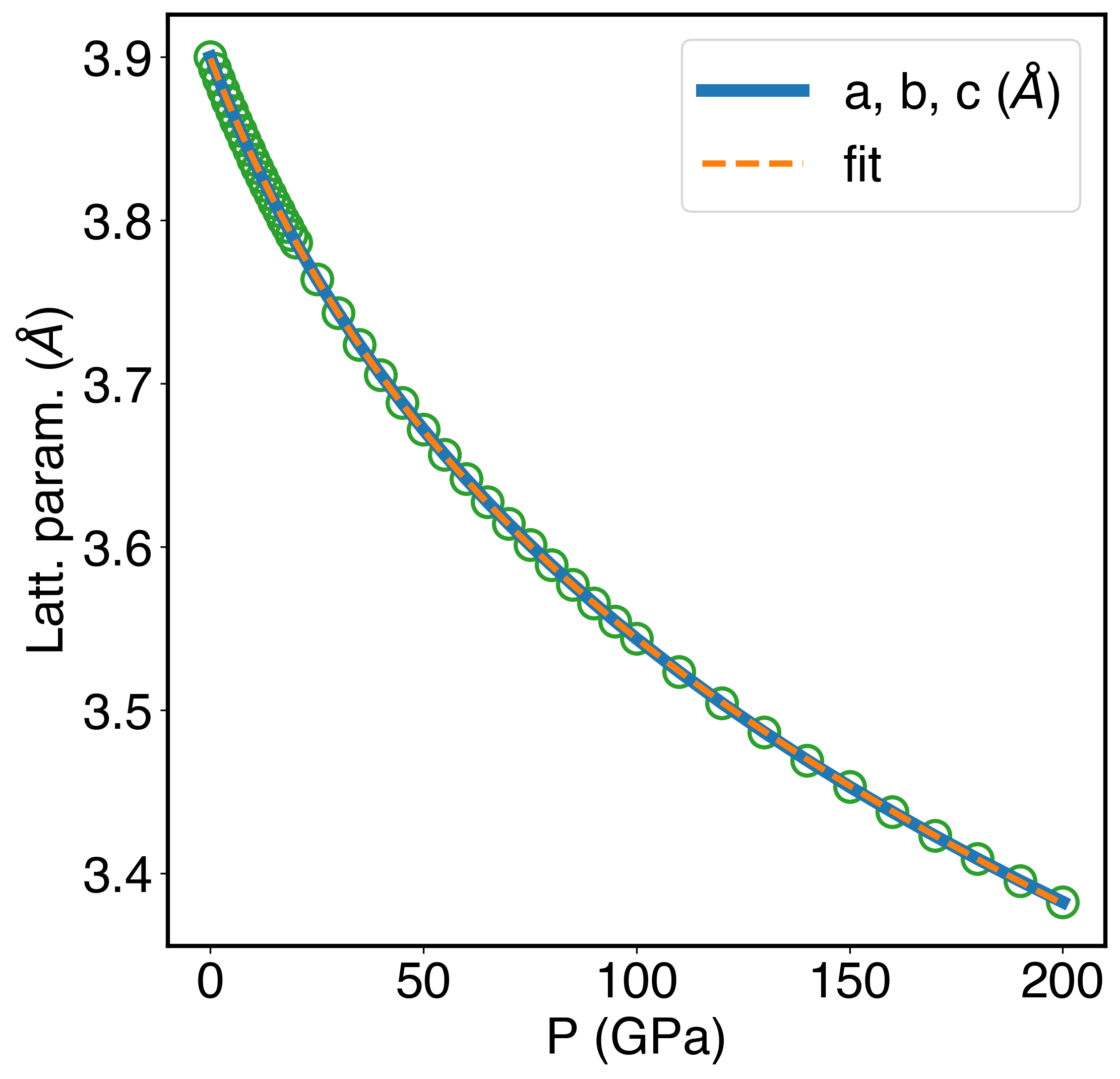}
	\caption{Lattice parameter of STO as a function of pressure from our DFT calculations.} 
	\label{fig:STO_lattpar}
\end{figure}

In this section, we report the main results for the structural relaxation of the SrTiO$_3$  (STO) substrate. In Fig. \ref{fig:STO_eos} we show the equation of states along with the results from a fit with the Birch-Murnaghan equation for STO as a function of pressure. In Fig. \ref{fig:STO_lattpar} we show the corresponding lattice parameter. 

The effect of isotropic pressure was computed in two steps, including the effect of the  SrTiO$_3$ (STO) substrate on the in-plane lattice parameter, as well as the effect of pressure on the $c$ axis. This strategy differs substantially from Ref. \cite{Christiansson_PRB_2023_PrSrNiO2_pressure}, where the in-plane lattice constant is kept constant, as it is not suited to study the rather isotropic pressures that are realized in experiments using diamond anvil cells.

\begin{figure}[htb]
	\centering
	\includegraphics[width=0.55\columnwidth]{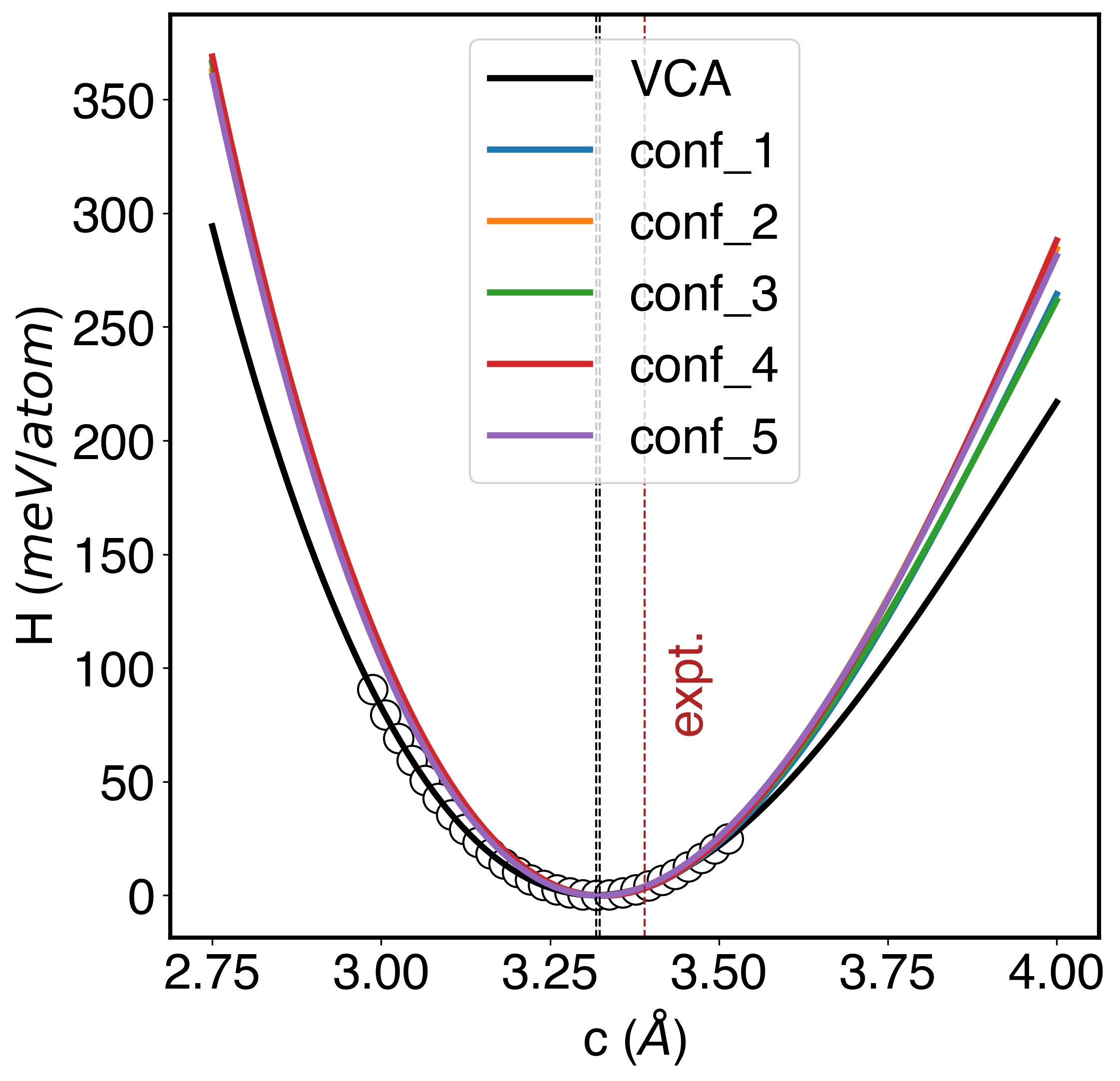}
	\caption{Enthalpy versus value of the $c$ axis for Pr$_{0.75}$Sr$_{0.25}$NiO$_2$ using the VCA and in five different supercells. The VCA value is shown as a black line, and individual points are shown as black dots. The results for the supercells are shown as colored lines.}
	\label{fig:prsrnio2_vca_test}
\end{figure}

Our method of computing the crystal structure of Pr$_{1-x}$Sr$_{x}$NiO$_2$, on the other hand, is motivated by the consideration of how the actual crystal is grown in experiments. Infinite-layer nickelates are grown over a perovskite substrate. Often STO is used \cite{Li_Nature_2019_supernickel,Osada_NanoLett_2020_PrNiO2,Li_PRL_2020_dome_nickel, Lee2020, Osada_AdvMat_2021_LaSrNiO2},
this also includes the experiments by Wang {\em et al.}~\cite{Wang_NatComm_2022_press_PrSrNiO2} which motivated the present study.
But, NdGaO$_3$ (NGO) \cite{Ikeda_PhysC_2014_ngo} and  (LaAlO$_3$)$_{0.3}$(Sr$_2$TaAlO$_6$)$_{0.7}$ (LSAT) \cite{lee2022character}  was employed as well. 
The nickelate layer is grown with a thickness between 10 and 100 nm \cite{Ikeda_PhysC_2014_ngo, Li_Nature_2019_supernickel, Osada_NanoLett_2020_PrNiO2, Lee2020, Li_PRL_2020_dome_nickel, Osada_AdvMat_2021_LaSrNiO2,Zeng_SciAdv_2021_super_CaNiO2} over a bulk substrate which can be regarded as infinite, and capped again with a few layers of the same substrate. Hence we work in the hypotheses that:
\begin{enumerate}
	\item The nickelate is forced to assume the same in-plane lattice constant of the substrate on which it is grown.
	\item In the $xy$ plane the elastic response of the system is dominated by the substrate, due to the substrate being much thicker.
	\item Along the $z$ direction, the nickelate is not constrained by the substrate, hence its response is independent from it.
\end{enumerate}
The above is then modelled in the two following steps:
\begin{enumerate}
	\item We compute the equation of state for bulk STO in the cubic phase. At a given pressure, the equation of state is used to extract the in-plane lattice constants $a = b$ (Supplementary Figure \ref{fig:STO_eos}).
	\item With $a$ and $b$ fixed to the value given by STO at the chosen pressure, we compute the enthalpy of the nickelate phase as a function of the $c$ axis, see Fig.~\ref{fig:prsrnio2_vca_test}. The $c$ value that minimizes the enthalpy corresponds to the equilibrium value. 
\end{enumerate}

\label{subsect:struct_relax}

\FloatBarrier
%\newpage

\subsection{Wannier Functions}
\label{subsect:wannier_results}
The DFT bandstructure calculated in VASP  is subsequently mapped  onto a 10- and a 1-band
Wannier basis using maximally localized Wannier orbitals and the  \verb|wannier90|  code \cite{mostofi2008wannier90}. Here, $d$ orbitals centered on Pr and Ni were used for the initial projections for the 10-bands calculations, and a single d$_{x^2-y^2}$ orbital centered on Ni was used as initial projection for the 1-band calculation. The reader interested in the energy ranges for the disentanglement and frozen windows of each calculation can find the corresponding input files in \cite{suppdata}. The Hamiltonian $H(\textbf{k})$ was obtained as the Fourier transform of the \verb|wannier90| output.

\begin{figure}[htb]
	\centering
	\includegraphics[width=0.49\columnwidth]{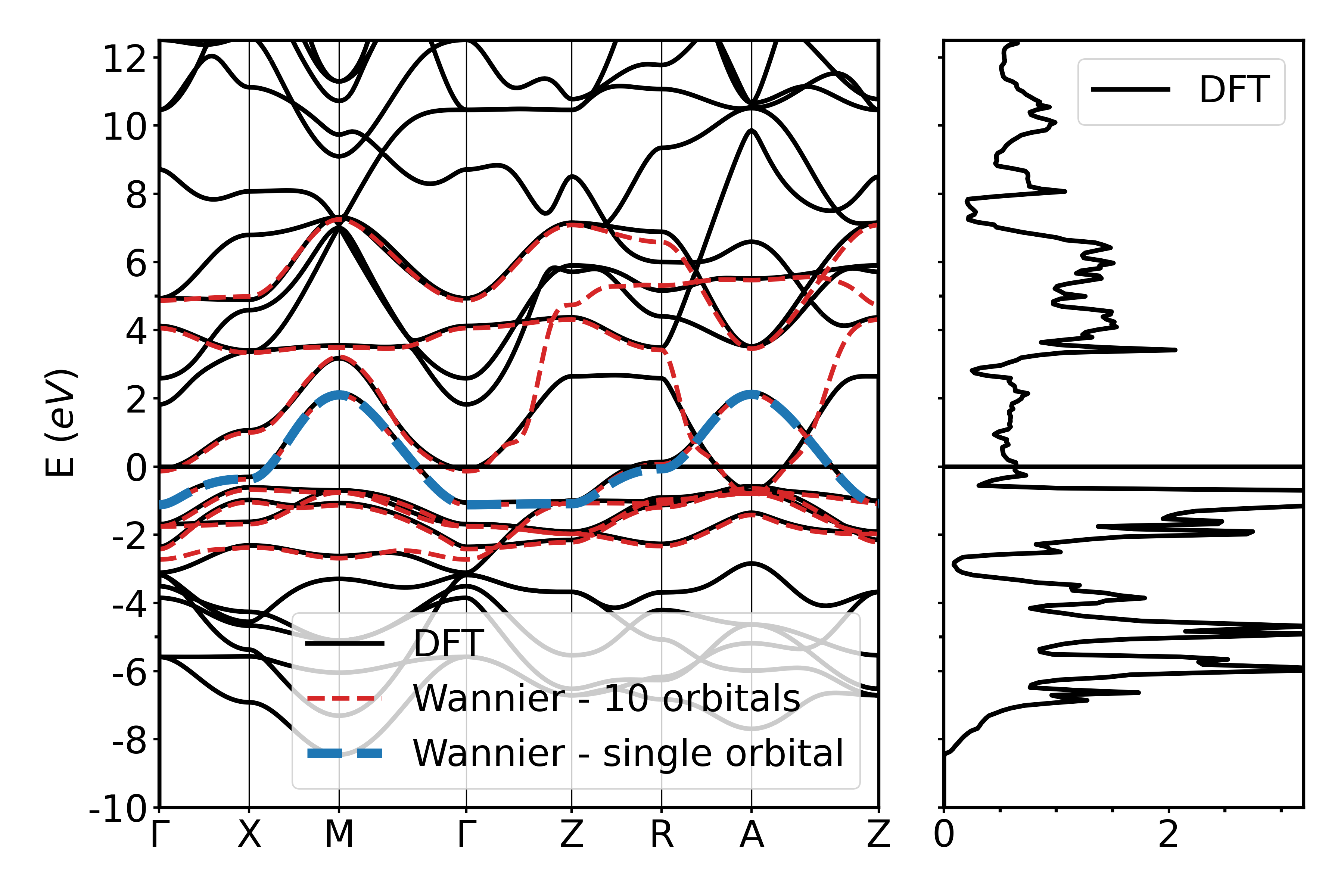}
	\includegraphics[width=0.49\columnwidth]{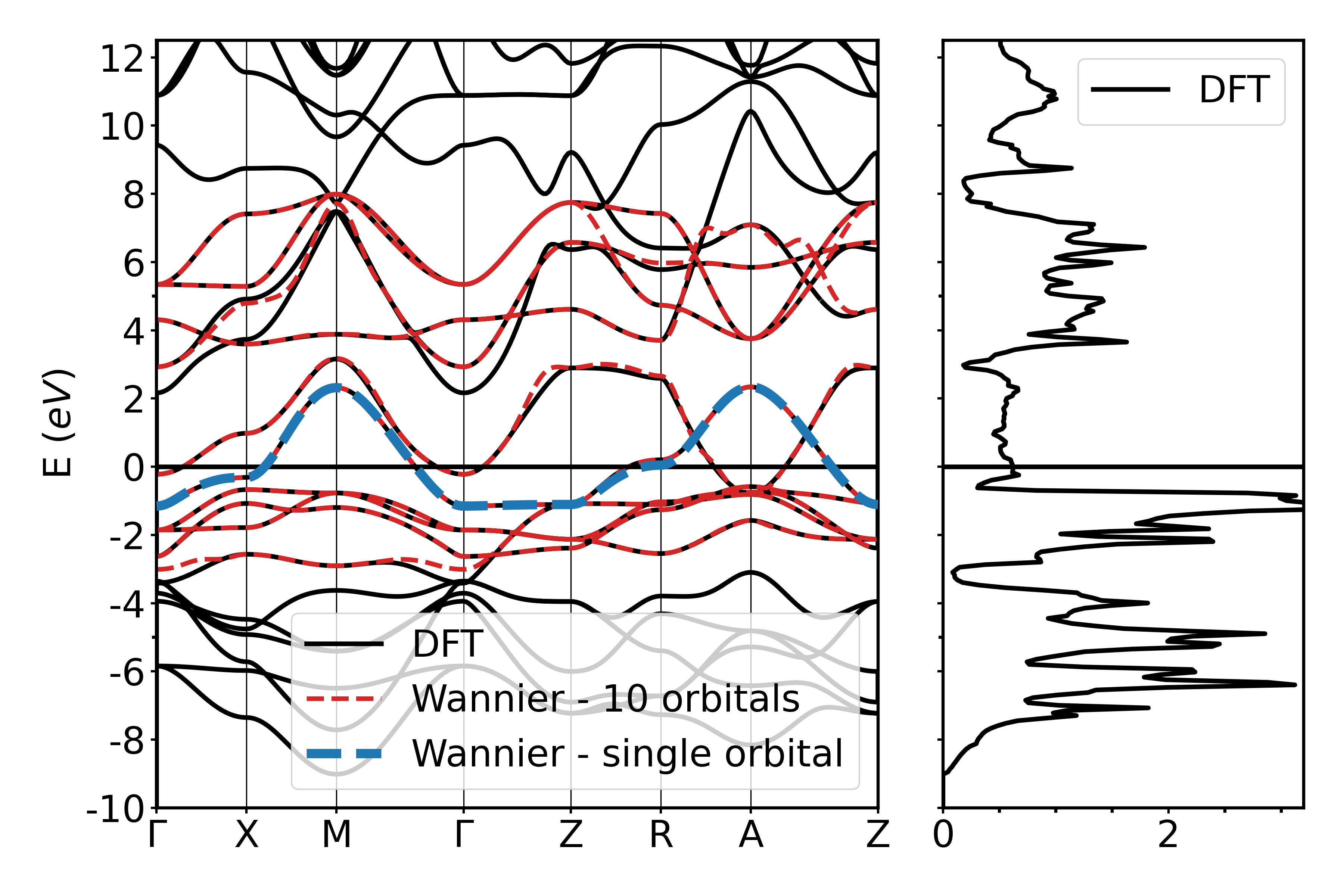}
	\includegraphics[width=0.49\columnwidth]{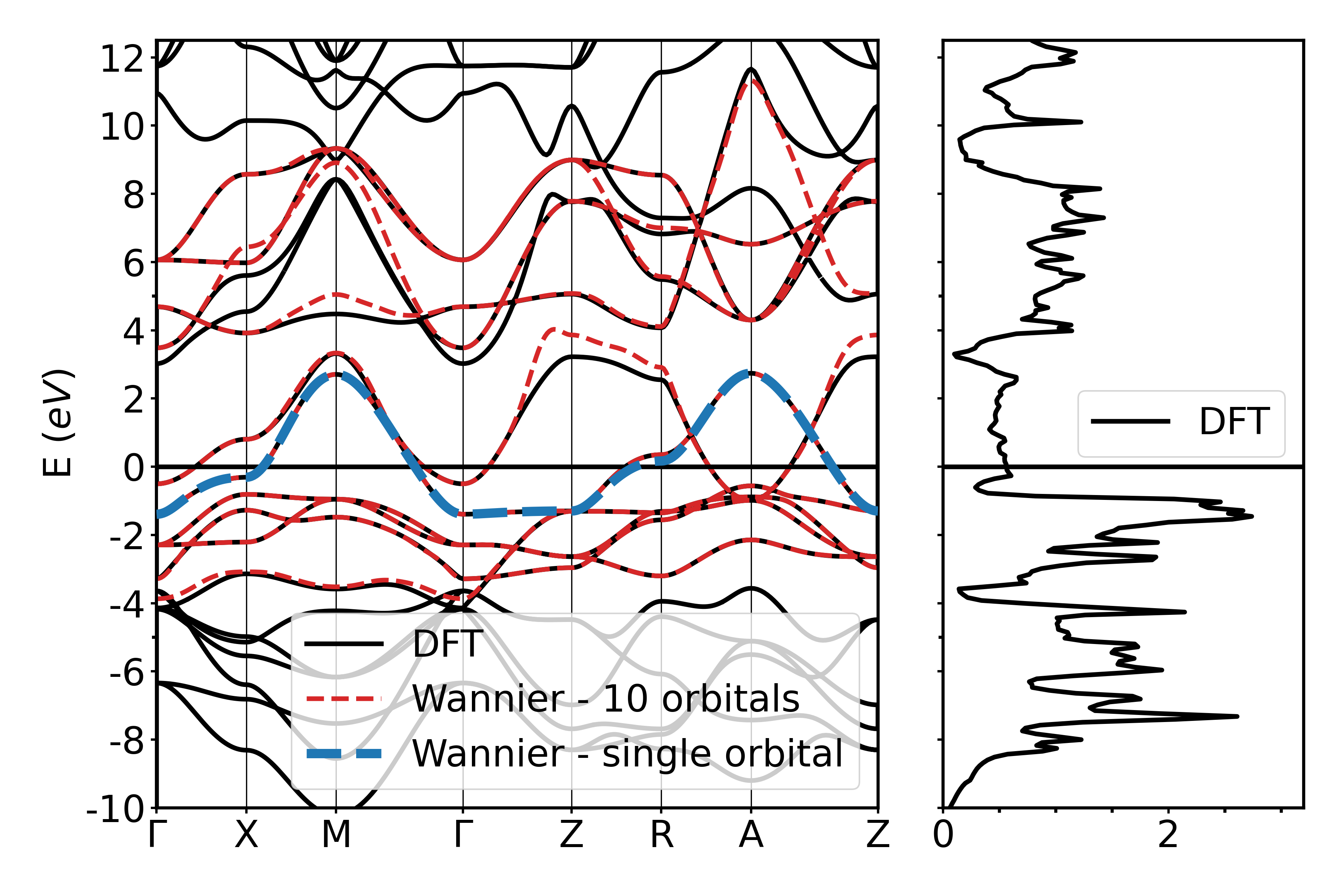}
	\includegraphics[width=0.49\columnwidth]{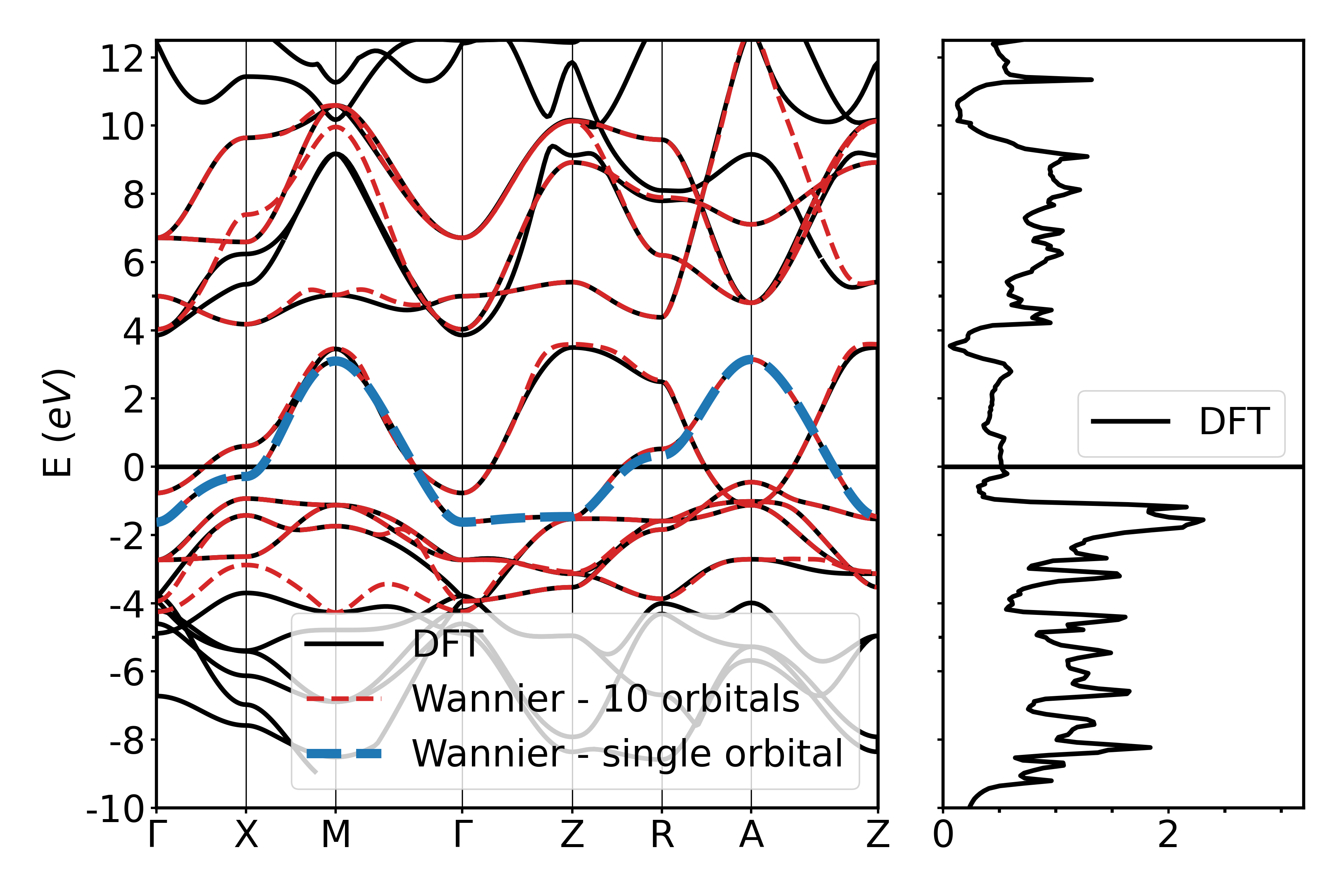}
	\includegraphics[width=0.49\columnwidth]{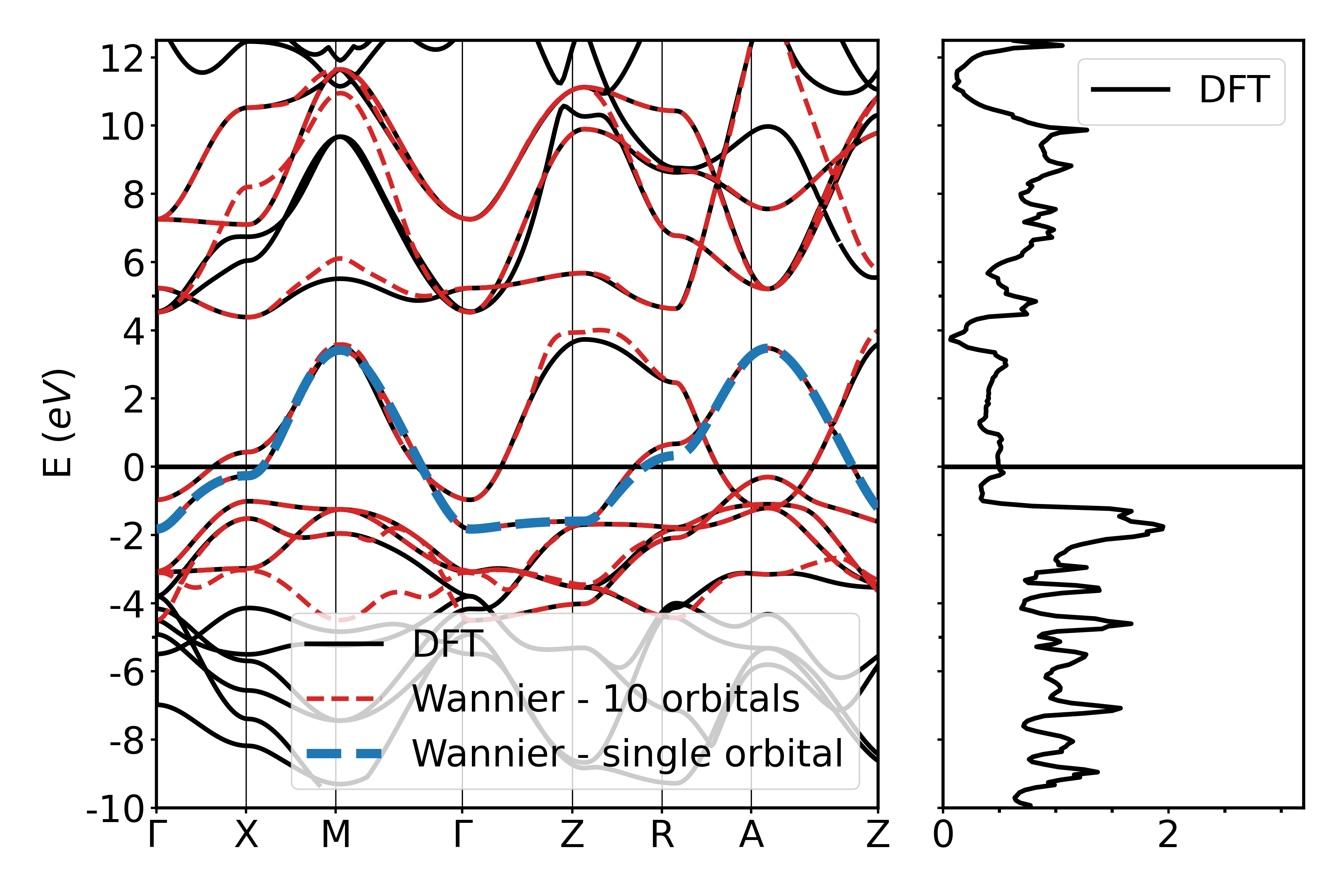}
	\caption{DFT electronic band structure and 10- and 1- band wannierization as a function of pressure. The DFT bands, the 10 bands wannierization and the 1 band wannierization are shown as black solid, red dashed, and blue dashed lines, respectively. The figures are shown in order of ascending pressure. From top left to bottom right: 0, 12, 50, 100, and 150 GPa, respectively.}
	\label{fig:wannier_pressure}
\end{figure}

Fig.~\ref{fig:wannier_pressure} shows
the direct comparison between Wannier and DFT bands.
While there are some deviations of the 10-band model and the DFT band above the Fermi energy where additional bands --besides the 10-bands considered-- cross, the agreement at low energies is excellent. And this is the relevant region for the subsequent DMFT calculation.

From the wannierization, we can also extract the hopping amplitudes $t$, $t'$, and $t''$ for nearest, next-nearest, and next-next-nearest neighbour hopping for the effective 1-orbital model. These
parameters 
for different pressures and dopings are reported in Table~\ref{tab:summary_dmft_inputs} and serve as a DMFT input for the 1-band calculation. In case of the fully-fledged 10-band DMFT calculation, we used the full  $H(\textbf{k})$ of the Wannier bands, without restriction to shorter-range hoppings. 

\begin{table}[htb]
	\centering
	\begin{tabular}{|c|c|c|c|c|c|c|c|c|}
		\hline
		Pressure   &  Doping  &  $t$  &   $t'$  &   $t''$  &  U   &    U   &  n$_{eff}$ Ni $d_{x^2-y^2}$\\
		(GPa)     &         & (eV)    &   (eV) &   (eV)  & (eV) &   ($t$) &  ($e/band$) \\
		\hline
		\hline
		0         &   0\%   & -0.388  &   0.097 & -0.049  &   3.4 &   8.77  & 0.973 \\ 
		12         &   0\%   & -0.416  &   0.099 & -0.052  &   3.4 &   8.17  & 0.965 \\
		50         &   0\%   & -0.483  &   0.109 & -0.059  &   3.4 &   7.04  & 0.933 \\
		100         &   0\%   & -0.558  &   0.112 & -0.067  &   3.4 &   6.10  & 0.887 \\
		150         &   0\%   & -0.617  &   0.118 & -0.073  &   3.4 &   5.51  & 0.858 \\
		\hline
		0         &  18\%   & -0.388  &   0.097 & -0.049  &   3.4 &   8.77  & 0.838 \\ 
		12         &  18\%   & -0.416  &   0.099 & -0.052  &   3.4 &   8.17  & 0.831 \\
		50         &  18\%   & -0.483  &   0.109 & -0.059  &   3.4 &   7.04  & 0.817 \\
		100         &  18\%   & -0.558  &   0.112 & -0.067  &   3.4 &   6.10  & 0.775 \\
		150         &  18\%   & -0.617  &   0.118 & -0.073  &   3.4 &   5.51  & 0.742 \\
		\hline
		0         &   0\%   & -0.388  &   0.097 & -0.049  &   3.4 &   8.77  & 0.973 \\ 
		0         &  10\%   & -0.388  &   0.097 & -0.049  &   3.4 &   8.77  & 0.902 \\
		0         &  15\%   & -0.388  &   0.097 & -0.049  &   3.4 &   8.77  & 0.862 \\
		0         &  20\%   & -0.388  &   0.097 & -0.049  &   3.4 &   8.77  & 0.823 \\
		0         &  25\%   & -0.388  &   0.097 & -0.049  &   3.4 &   8.77  & 0.790 \\
		0         &  30\%   & -0.388  &   0.097 & -0.049  &   3.4 &   8.77  & 0.767 \\
		\hline
		50         &   0\%   & -0.483  &   0.109 & -0.059  &   3.4 &   7.04  & 0.933 \\ 
		50         &   5\%   & -0.483  &   0.109 & -0.059  &   3.4 &   7.04  & 0.905 \\
		50         &  10\%   & -0.483  &   0.109 & -0.059  &   3.4 &   7.04  & 0.872 \\
		50         &  15\%   & -0.483  &   0.109 & -0.059  &   3.4 &   7.04  & 0.840 \\
		50         &  20\%   & -0.483  &   0.109 & -0.059  &   3.4 &   7.04  & 0.802 \\
		50         &  25\%   & -0.483  &   0.109 & -0.059  &   3.4 &   7.04  & 0.762 \\
		50         &  30\%   & -0.483  &   0.109 & -0.059  &   3.4 &   7.04  & 0.722 \\
		50         &  40\%   & -0.483  &   0.109 & -0.059  &   3.4 &   7.04  & 0.643 \\
		\hline
	\end{tabular}
	\caption{Calculated quantities for the single-band Hubbard model of the Ni $d_{x^2-y^2}$ band at the pressures and physical Sr-doping values $x$ employed in this paper. The effective filling $n = 1-\delta$ is reported.}
	\label{tab:summary_dmft_inputs}
\end{table}
\FloatBarrier

\subsection{Constrained random phase approximation}
\label{subsect:methods_hubbard}
This Wannier Hamiltonian needs to be supplemented by the Coulomb interaction. To this end, the static Hubbard interaction $U$ was computed from first principles using the constrained random phase approximation (cRPA) in the Wannier basis \cite{miyake:085122}
for entangled band-structures \cite{miyake:155134}.
Here, the underlying electronic structure was computed from DFT using a full-potential linearized muffin-tin orbital (fplmto) method \cite{fplmto} and the local density approximation, applied to bulk LaNiO$_2$ using the relaxed tetragonal structures from Sec.~\ref{subsect:methods_relaxation}.
The calculations use $10^3$ reducible $\textbf{k}$-points, and a muffin-tin radius (RMT) for Ni of 1.97 Bohr radii, except for $P=100$ GPa, where RMT=1.9. At $P=50$ GPa the difference in RMT changes $U$ by merely 0.4\%.
The cRPA results for the 1-band model are shown in 
Table \ref{tab:U}.  
%\kh{have moved the table to the main text but would put the U values here}
Note that the Hubbard $U$ can indeed display a non-trivial dependence on pressure \cite{Upressure}. E.g., in cuprates in-plane compression increases the local interaction in a \dxsqysq{} setting \cite{Ivashko2019}. 

We find both the screened and the bare interaction---$U$ and $V$---to be essentially insensitive to pressure: Both the localization of the \dxsqysq-derived Wannier function and the screening remain constant.
This justifies keeping $U$ unchanged with pressure. To account for the frequency-dependence of $U$ in cRPA a slightly larger value is needed, and we use a fixed $U=3.4\,$eV as is stated in Table~\ref{tab:summary_dmft_inputs} was already employed in
\cite{Kitatani_npj_2020_renaissance}.

\begin{table}
	\centering
	\begin{tabular}{c|c|c|c|c|c}
		$P$ [GPA] &0 & 12.1 & 20 & 50 & 100    \\
		\hline
		$U$ [ev] & 2.88 &  2.88 & 2.88 & 2.96 & 2.93 \\
		$V$ [eV] & 19.4 & 19.3 & 19.2 & 19.3 & 19.0 
	\end{tabular}
	\caption{Local Hubbard interaction $U$ and bare (unscreened) Coulomb interaction $V$  in the maximally localized Wannier function basis for the  1-band (\nidxsqysq-band) model as calculated by cRPA.}
	\label{tab:U}
\end{table}

Given their weak pressure dependence, we also keep the interaction parameters fixed for the 10-band calculation. These have been calculated before in  cRPA~\cite{Si_PRL_2020_LaNiO2_H} and are
displayed in Table \ref{tab:10band_DMFT}. 

\begin{table}[htb]
	\centering
	\begin{tabular}{|c|c|c|c|c|c|c|c|}
		\hline
		Pressure  & Temperature &  U$_{Ni}$      &     J$_{Ni}$  &    U'$_{Ni}$  &  U$_{Nd}$      &     J$_{Nd}$  &    U'$_{Nd}$  \\
		(GPa)    & (K) &   (eV)   &   (eV) &   (eV)  &   (eV)   &   (eV) &   (eV)  \\
		\hline
		0        & 300 &  4.40     &   0.65 & 3.10   & 2.50     & 0.25   & 2.00 \\
		12       & 300 &  4.40     &   0.65 & 3.10   & 2.50     & 0.25   & 2.00 \\
		50       & 300 &  4.40     &   0.65 & 3.10   & 2.50     & 0.25   & 2.00 \\
		100      & 300 &  4.40     &   0.65 & 3.10   & 2.50     & 0.25   & 2.00 \\
		150      & 300 &  4.40     &   0.65 & 3.10   & 2.50     & 0.25   & 2.00 \\
		\hline
		\hline
	\end{tabular}
	\caption{Intra- and inter-orbital Coulomb repulsion (U, U') and Hund coupling J for the Kanamori Hamiltonian employed in the 10-band DMFT calculations.}
	\label{tab:10band_DMFT}
\end{table}

%We summarize how the three main parameters of the single-band Hubbard model, the local Coulomb interaction $U$, the hopping $t$ (and $t'$/$t''$), and the filling $n$, were computed for the \nidxsqysq{} orbital. The Hubbard interaction $U$ was computed within the constrained Random-Phase Approximation (cRPA) \cite{Takashi_PRB_2008_cRPA}. We find that from 0 to 50 GPa the bare Coulomb potential $V$ remains essentially constant. \note{Jan add details, also maybe the figure which shows that U is constant wr.t. pressure in the supplementary?}. The hoppings $t$/$t'$/$t''$ were extracted from the real-space Wannier Hamiltonian for \nidxsqysq{}. Their 

\subsection{Dynamical mean-field theory} 
\label{subsect:DMFT}
DMFT\cite{Georges1996,held2007electronic} calculations were performed using \verb|w2dynamics| version 1.1.3 \cite{Parragh_PRB_2012_w2dynamics, Wallerberger_CompPhysComm_2019_w2dynamics}. All the input files are available as extended data at Ref. \cite{suppdata}.

\paragraph{10-band case}
In the 10-band case, we employ a Kanamori Hamiltonian, and considered the Nd and Ni atoms as two different impurity sites, with interactions described in Supplementary Table \ref{tab:10band_DMFT}. The convergence of the local Green's function was achieved through a three-step process. The first two steps were performed with an increasing sampling of the quantum Monte-Carlo solver, for a total of 30 iterations, while a third, final step with a larger number of iterations was employed to better sample the Green's function.
\paragraph{1-band case}
In the 1-band case we employed a Hamiltonian with only density-density interaction and the same two-step scheme of the 10-band case, with the addition of a fourth step with much larger sampling to obtain the local two-particle Green's function.

\subsection{Dynamical vertex approximation}
\label{subsect:DGA}
Based on this 1-band model, we perform ladder D$\Gamma$A calculations to obtain the nonlocal magnetic and superconducting susceptibility starting from the local two-particle Green's function.
As explained in more details in \cite{Kitatani2022}, from the local two-particle Green's function, first  a local vertex $\Gamma$ that is irreducible in the particle-hole channel is determined. From the local $\Gamma$ in turn we calculate the D$\Gamma$A lattice susceptibility using the Moriya-$\lambda$ correction \cite{Toschi2007,Katanin2009,Rohringer2018}. In our case, the dominant susceptibility is the  magnetic one.
From this susceptibility in turn we extract the
irreducible vertex in the particle-particle or Cooper channel $\Gamma_{pp}$, cf.~\cite{Kitatani2022,Kitatani2019}. Finally, from $\Gamma_{pp}$ and the bare susceptibility, we obtain the superconducting eigenvalue $\lambda$. Superconductivity is signalled by the leading eigenvalue (in our case this is in the $d$-wave channel) approaching one.

Following Ref.~\cite{Kitatani_npj_2020_renaissance}, where low critical temperatures did not allow for a direct calculation below the critical temperature, the superconducting eigenvalue $\lambda$ was fitted with the function $\lambda(\beta) = A - B*np.log(\beta)$ to extrapolate the result to $\lambda = 1$.

\FloatBarrier
\newpage
\section{Additional results}
\label{SSec:results}
In this section, we report additional results from our study.

\subsection{DFT+DMFT}
\label{subsec:results:DMFT}
\begin{figure}[htb]
	\centering
	\includegraphics[width=0.48\columnwidth]{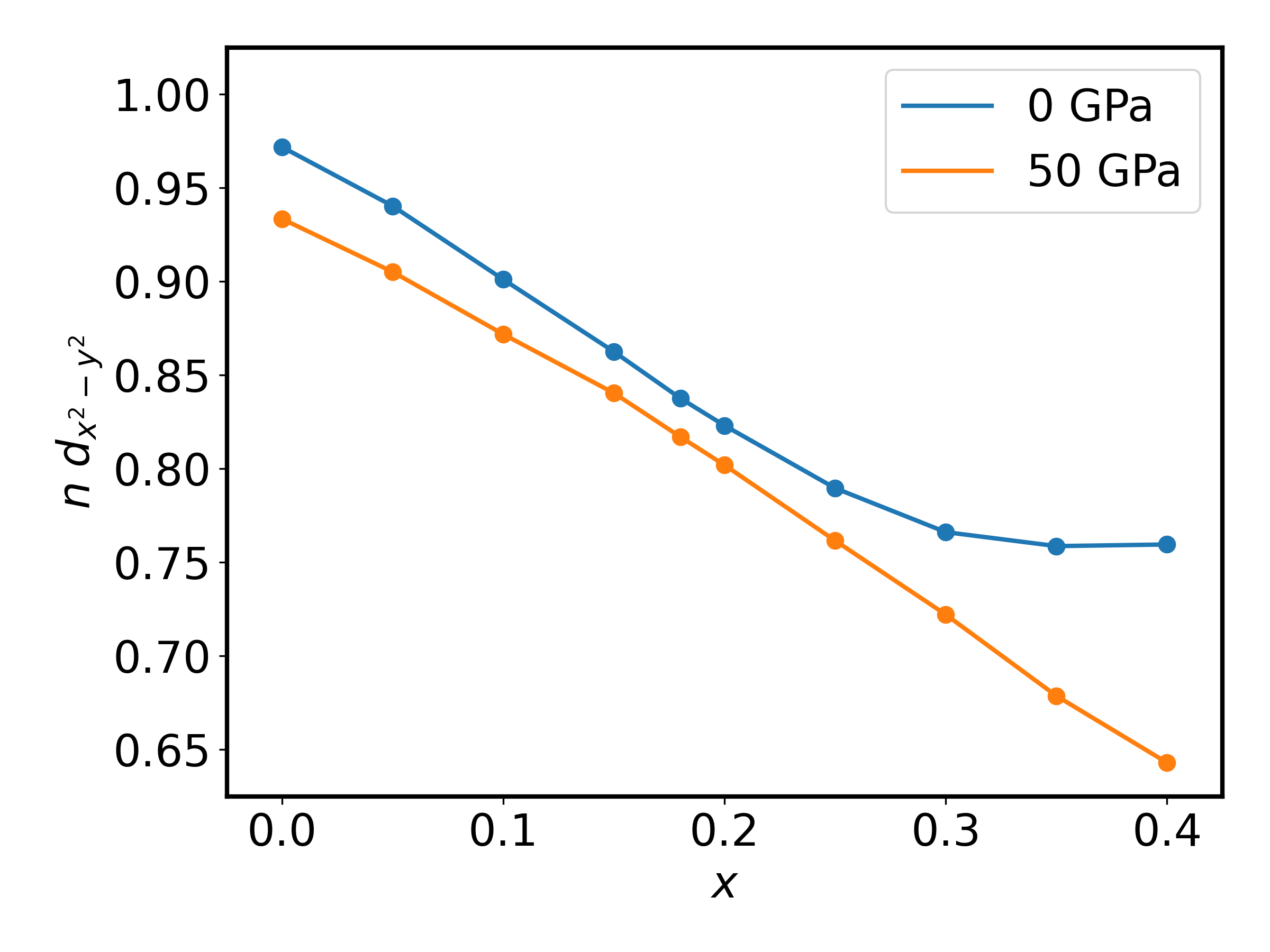}
	\includegraphics[width=0.48\columnwidth]{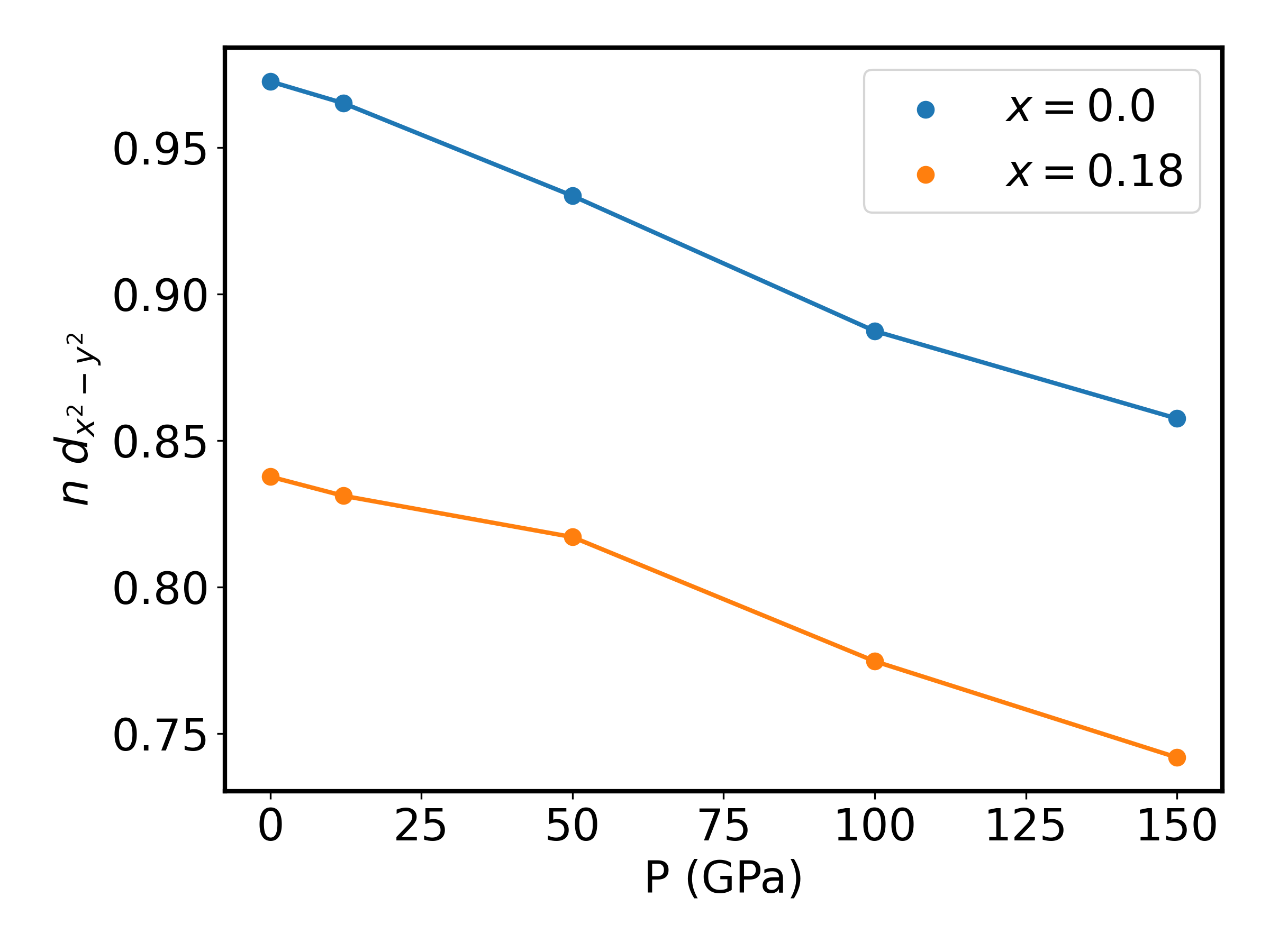}
	\caption{{Left panel}: Effective filling $n_{d_{x^2-y ^2}}$ of the \nidxsqysq band as a function of Sr doping $x$, calculated in DMFT for the 10-band model. The results for 0 and 50 GPa are indicated as blue and orange lines, respectively. 
		{Right panel}:  $n_{d_{x^2-y ^2}}$ of the \nidxsqysq band as a function of pressure at fixed doping $x = 0.00$ (blue curve) and $x = 0.18$ (orange curve). The hole doping of the main text is $\delta=1- n_{d_{x^2-y^2}}$.}
	\label{fig:pressure_eff_filling}
\end{figure}

In Fig. \ref{fig:pressure_eff_filling} we show the effective filling of the \nidxsqysq band, obtained from the 10-band DMFT calculations as described in the main text. We computed the doping dependency at 0 and 50 GPa, and the pressure dependence at fixed Sr concentration $x = 0.18$. 
In general, the filling of the \nidxsqysq band decreases linearly with increasing pressure and with larger Sr concentration. However, at 0 GPa the decrease flattens out for $x \geq 0.3$, while it remains linear at 50 GPa.

\begin{figure*}[htb]
	\centering
	\includegraphics[width=0.95\columnwidth]{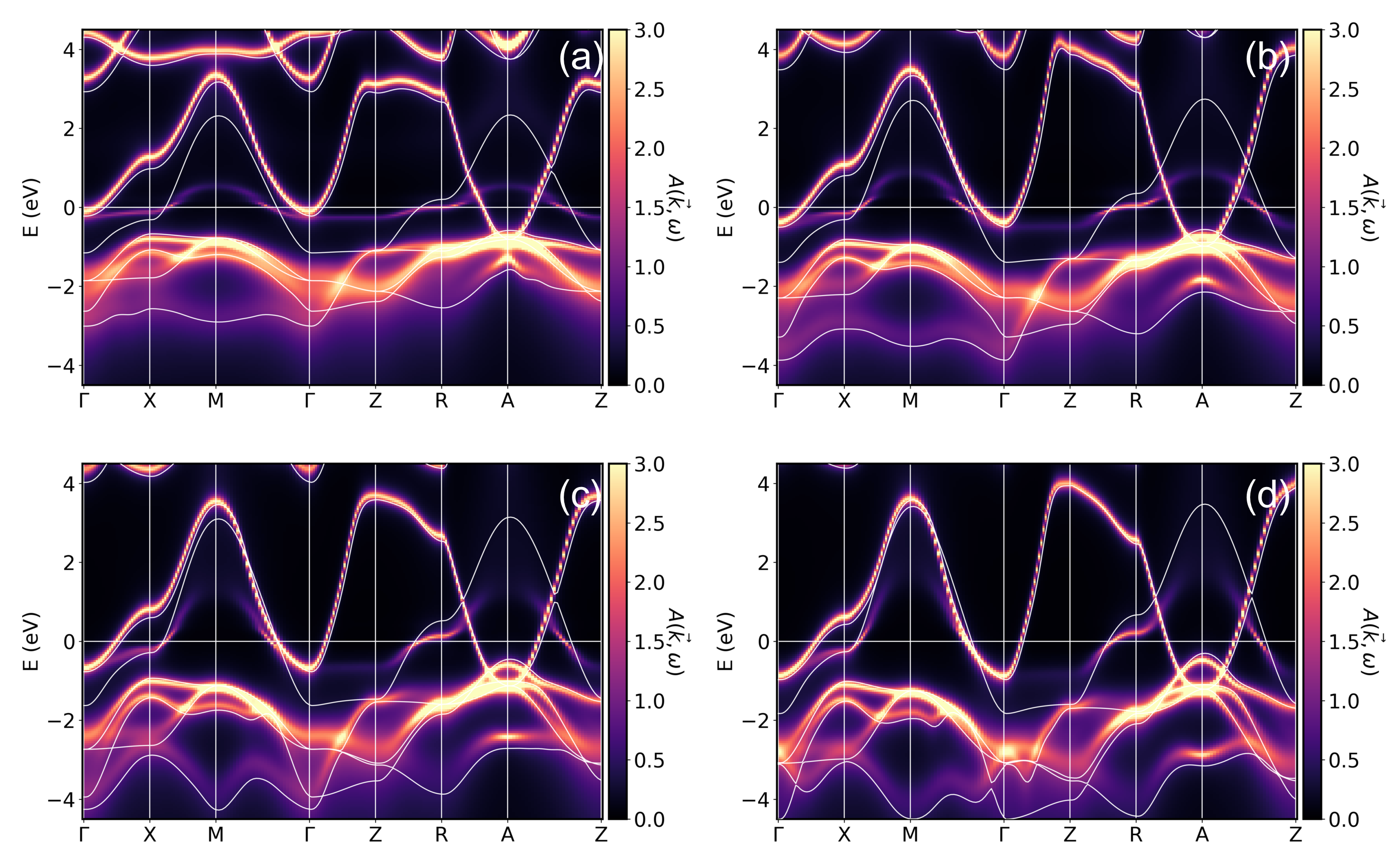}
	\caption{Evolution of the spectral function $A(\textbf{k}, \omega)$ of PrNiO$_2$  along a high-symmetry path in reciprocal space for the 10-bands DMFT calculations. (a), (b), (c), and (d) correspond to calculations at 12, 50, 100, and 150 GPa, respectively. The spectral function $A(\textbf{k}, \omega)$ is shown as a color scale. The DFT bands interpolated from Wannier functions are shown as white lines.}
	\label{fig:10_bands_0_doping_vs_pressure}
\end{figure*}

\begin{figure*}[htb]
	\centering
	\includegraphics[width=0.75\columnwidth]{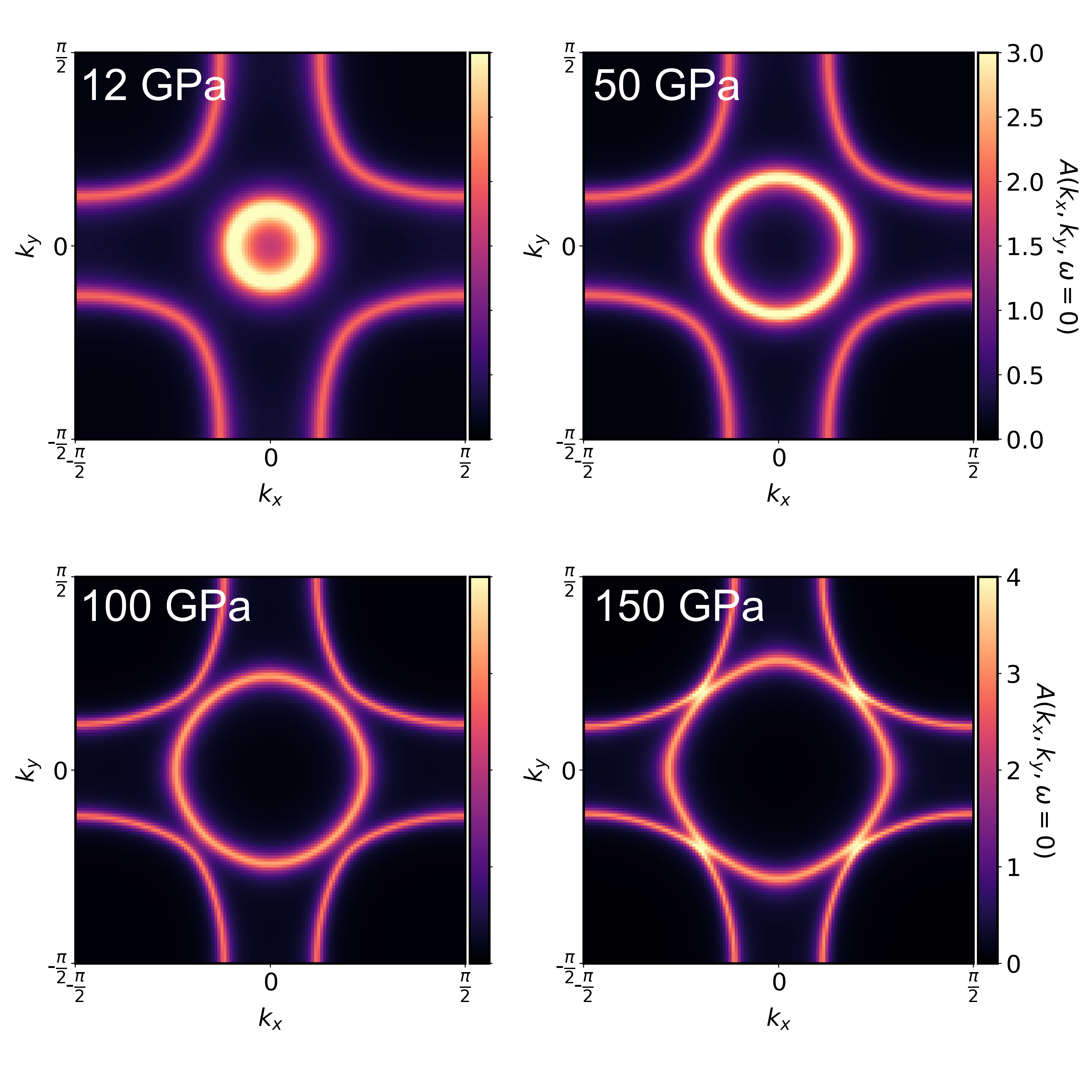}
	\caption{Evolution of the spectral function $A(\textbf{k}, \omega)$  of PrNiO$_2$  along the Fermi surface at $k_{z} = 0$ for the undoped PrNiO$_2$ as a function of pressure.}
	\label{fig:fs_pressure}
\end{figure*}

In addition to Fig.~2 of the main text, we show in Fig~\ref{fig:10_bands_0_doping_vs_pressure} the DMFT bandstructure of the parent compound, PrNiO$-2$, at further pressures,  visualizing the evolution with pressure. This is supplemented by the Fermi surface displayed in Fig.~\ref{fig:fs_pressure} for the same pressure and doping $x=0$. Note that at 150\,GPa the
electron pocket around $\Gamma$ becomes so large that it touches the \nidxsqysq band.  At pressures higher than this point, the description of the correlated system in terms of 1 correlated orbital plus decoupled pockets likely needs to be refined.

\begin{figure}[htb]
	\centering
	\includegraphics[width=0.48\columnwidth]{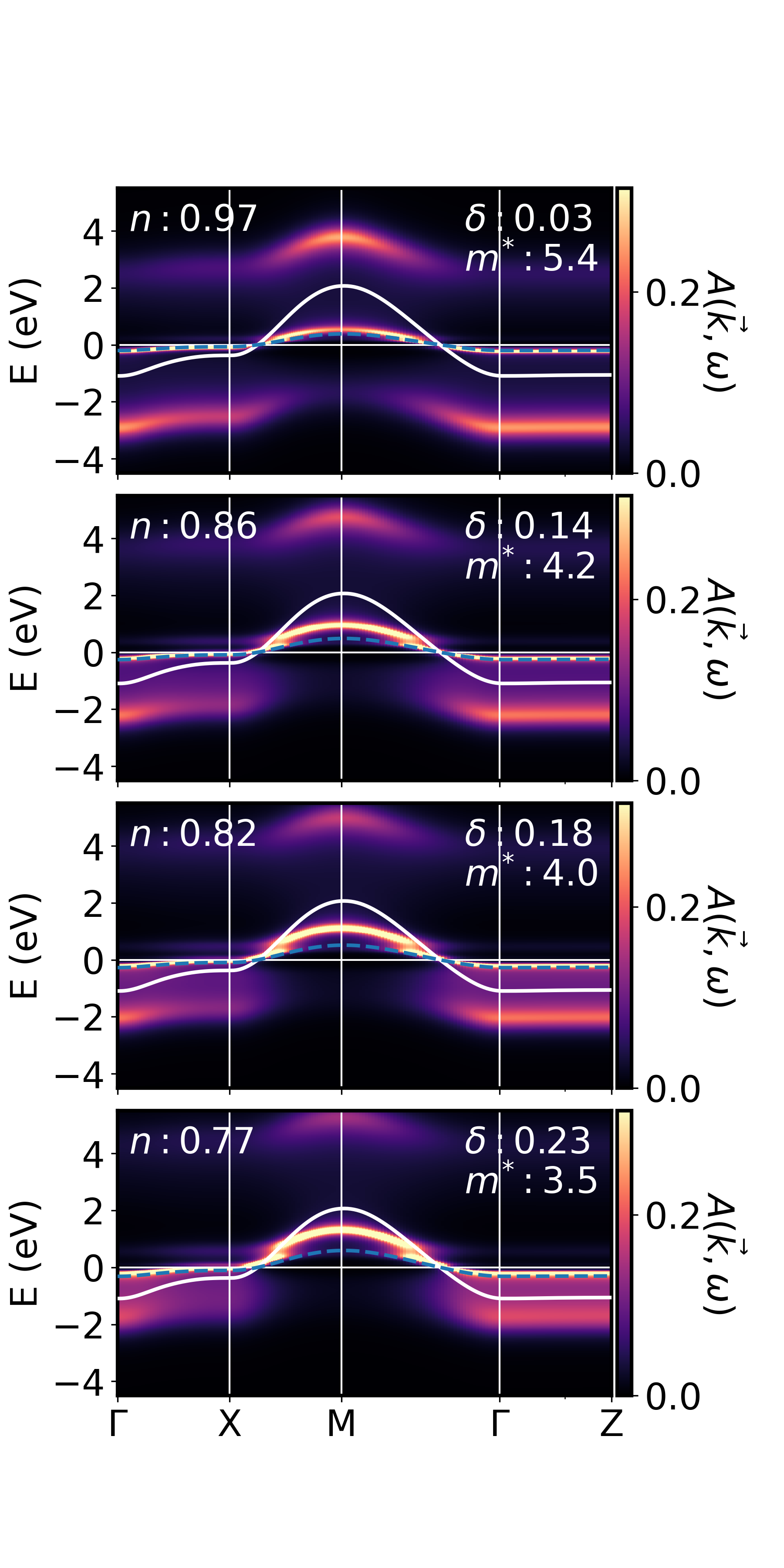}
	\caption{Spectral function computed from the local self-energy at fixed pressure P = 0 GPa as a function of doping $x$, from 0.0 to 0.30, at $\beta = 80 t$. The spectral function $A(\textbf{k}, \omega)$ is shown as a color gradient.}
	\label{fig:spfun_doping_0_GPa}
\end{figure}

\begin{figure}[htb]
	\centering
	\includegraphics[width=0.48\columnwidth]{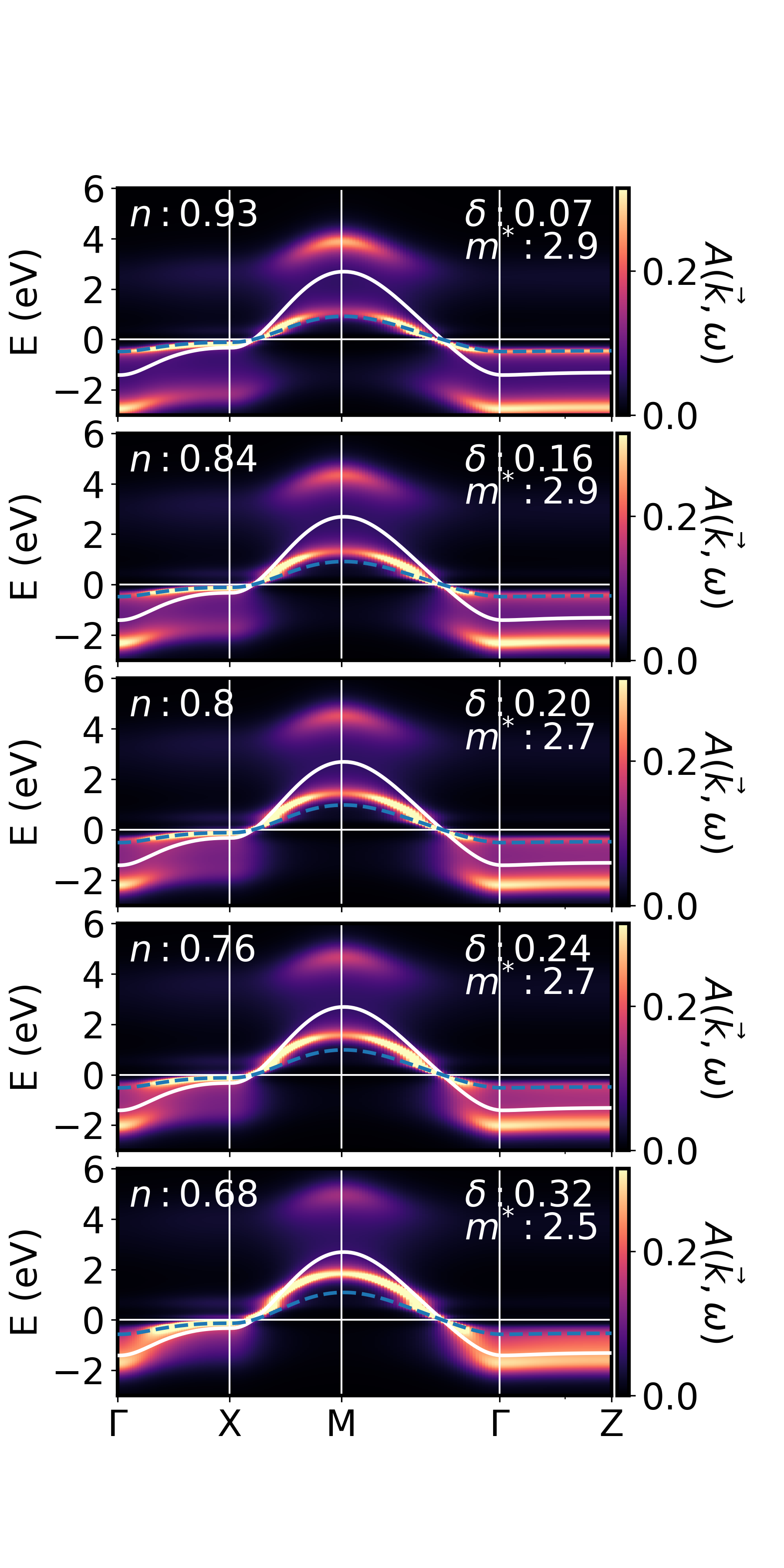}
	\caption{Spectral function computed from the local self-energy at fixed pressure P = 50 GPa as a function of doping $x$, from 0.0 to 0.30, at $\beta = 80 t$. The spectral function $A(\textbf{k}, \omega)$ is shown as a color gradient.}
	\label{fig:spfun_doping_50_GPa}
\end{figure}

\begin{figure}[htb]
	\centering
	\includegraphics[width=0.48\columnwidth]{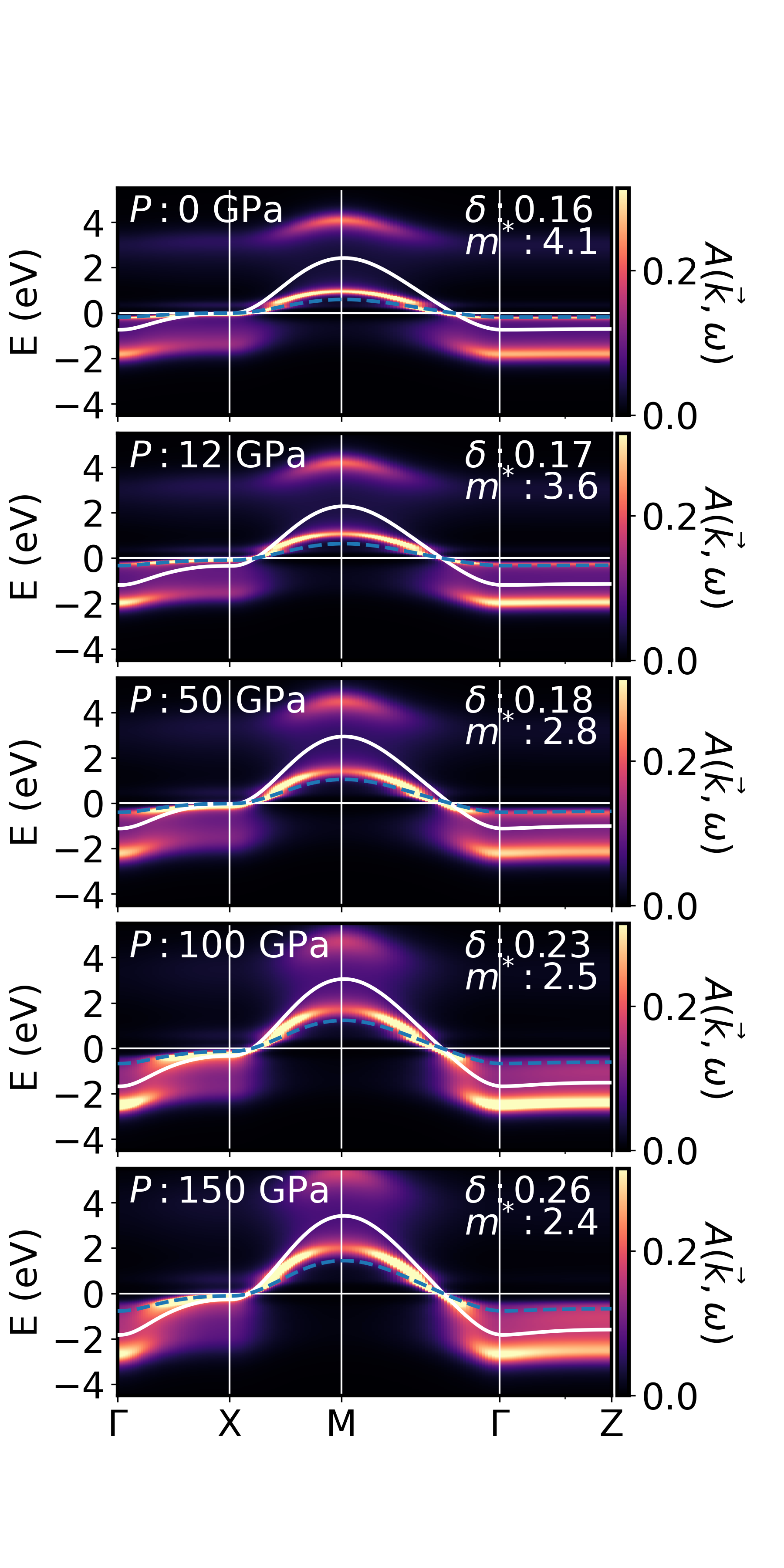}
	\caption{Spectral function computed from the local self-energy at fixed doping $x = 0.18$ and as a function of pressure, from 0 to 150, at $\beta = 80 t$. The spectral function $A(\textbf{k}, \omega)$ is shown as a color gradient.}
	\label{fig:spfun_d_0.18_pressure}
\end{figure}

In Fig.~3 of the main text, we showed the evolution of the 1-band DFT+DMFT spectrum of the parent compound ($x=0$) as a function of pressure. This corresponds to path (c) in Fig.~1 and 4 of  main text. Here, we also show the evolution with doping at 0\,GPa  [Fig.~\ref{fig:spfun_doping_0_GPa}, path (a)]
and 50\,GPa  [Fig.~\ref{fig:spfun_doping_50_GPa}, path (b)]; as well as the pressure dependence for doping $x=0.18$  [Fig.~\ref{fig:spfun_d_0.18_pressure}, path (d)].


\FloatBarrier
\newpage

\subsection{Superconductivity in \dga}
\label{subsec:results:DGA}
\begin{figure}[htb]
	\centering
	\includegraphics[width=0.48\columnwidth]{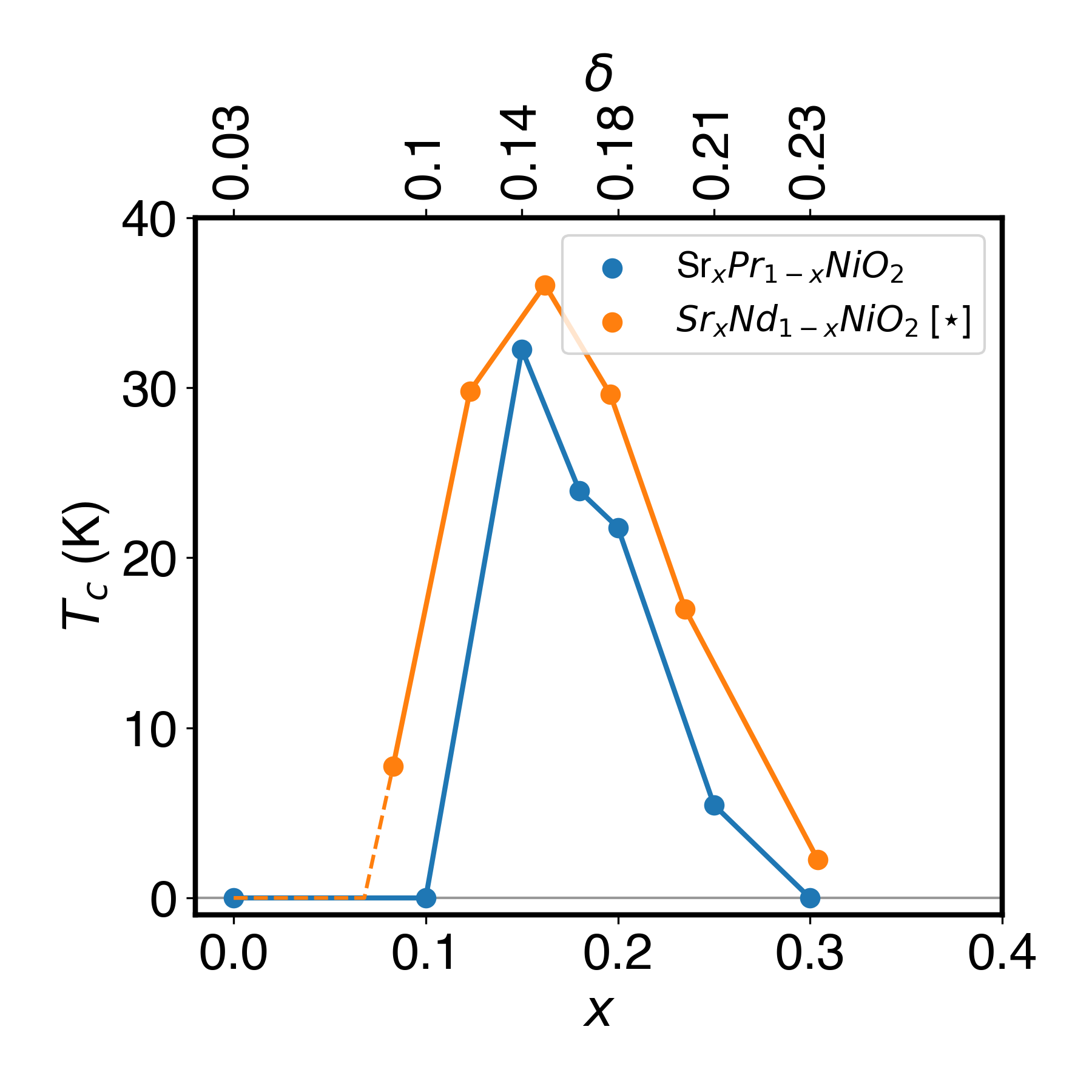}
	\caption{Comparison of the superconducting dome at 0 GPa calculated for Sr$_{x}$Pr$_{1-x}$NiO$_2$ (our work) and for Sr$_{x}$Nd$_{1-x}$NiO$_2$ (Ref. \cite{Kitatani_npj_2020_renaissance}).}
	\label{fig:tc_vs_Nd}
\end{figure}

In this section we present some additional comparisons of the superconducting phase diagram.
First, in Fig.~\ref{fig:tc_vs_Nd}, we compare the superconducting dome of  Sr$_{x}$Pr$_{1-x}$NiO$_2$ to that of  Sr$_{x}$Nd$_{1-x}$NiO$_2$ calculated in \cite{Kitatani_npj_2020_renaissance}. Both phase diagrams are in good agreement and the deviations, including the slightly lower \Tc in  Sr$_{x}$Pr$_{1-x}$NiO$_2$ can be explained by the different $A$ cation (Pr instead of Nd). 

Second in Fig.~\ref{fig:tc_and_t_p}, we plot the change of \Tc{} as a function of pressure together with the change of $t$. This comparison clearly reveals that the difference  between the parent compound (left) and 18\% Sr-doping (right) originates from the parent compound moving to optimal doping at 100\,GPa, whereas the doped sample moves from optimal doping to overdoped with pressure. 

\begin{figure}[htb]
	\centering
	\includegraphics[width=0.75\columnwidth]{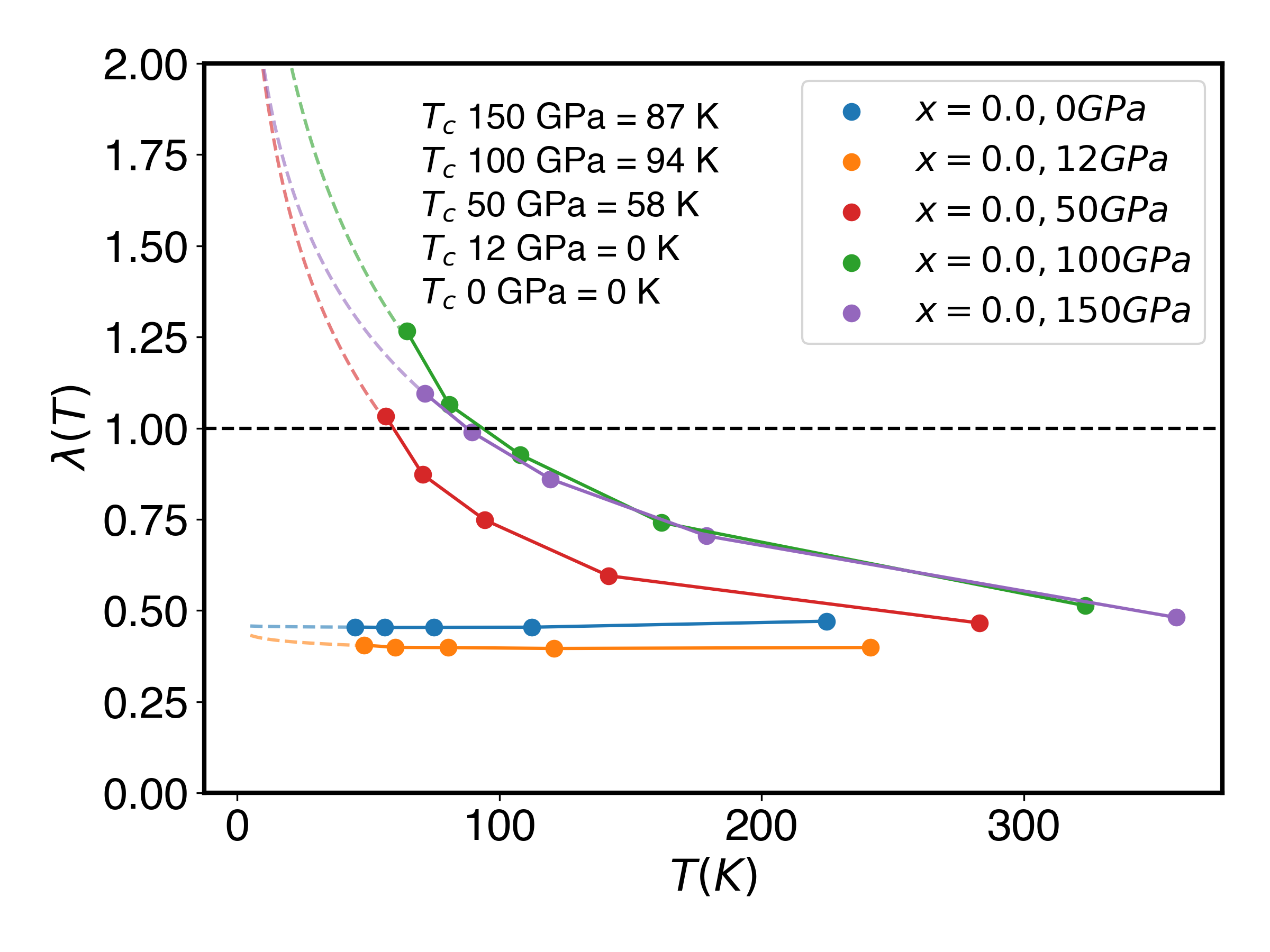}
	\caption{Leading superconducting eigenvalue $\lambda$ as a function of temperature at $x = 0$ and as a function of pressure. The critical temperature is selected as the value at which the curve (or its extrapolation) crosses the $\lambda = 1$ line.}
	\label{fig:lambda_vs_T_PrNiO2_pressure}
\end{figure}

\begin{figure}[htb]
	\centering
	\includegraphics[width=0.49\columnwidth]{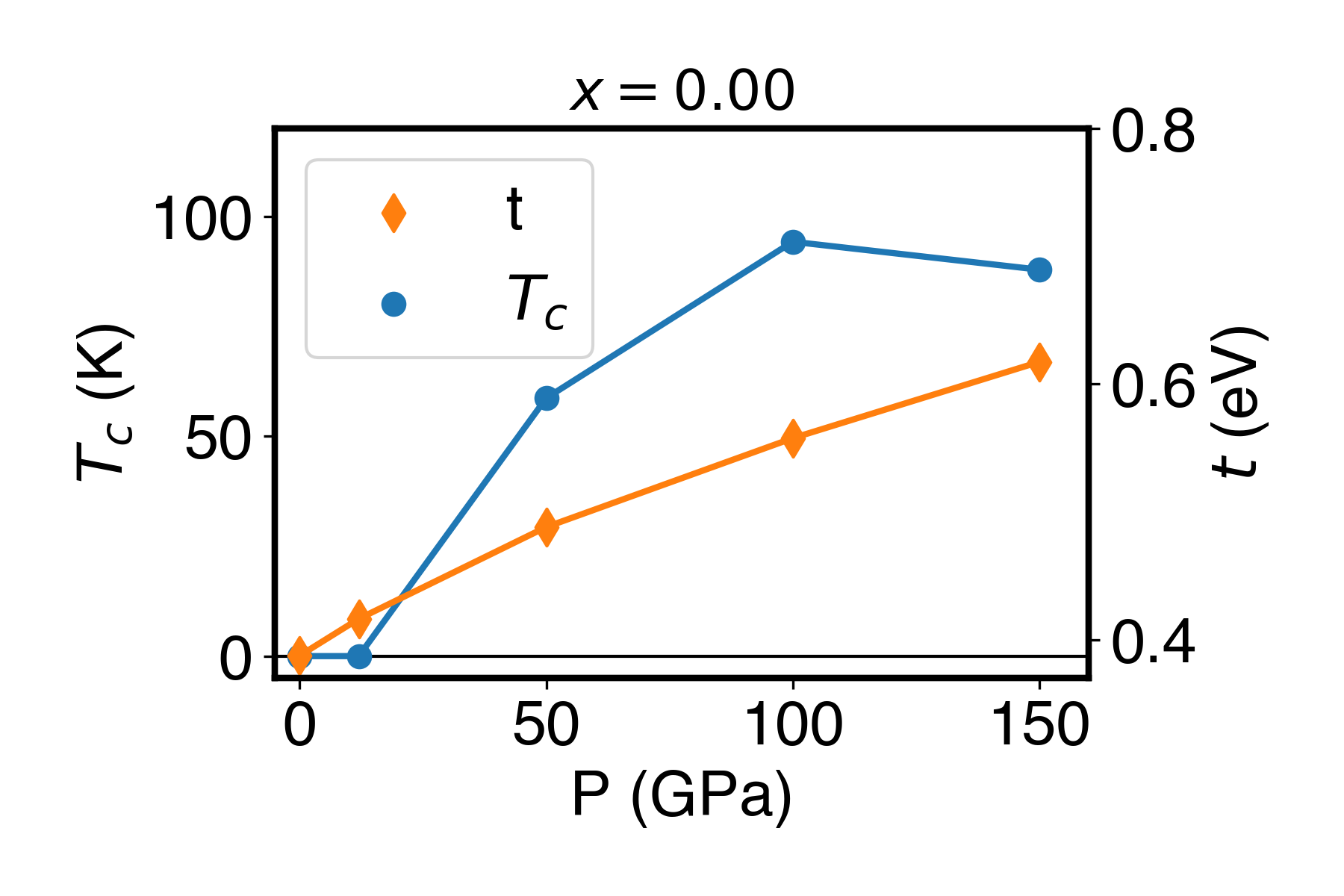}
	\includegraphics[width=0.49\columnwidth]{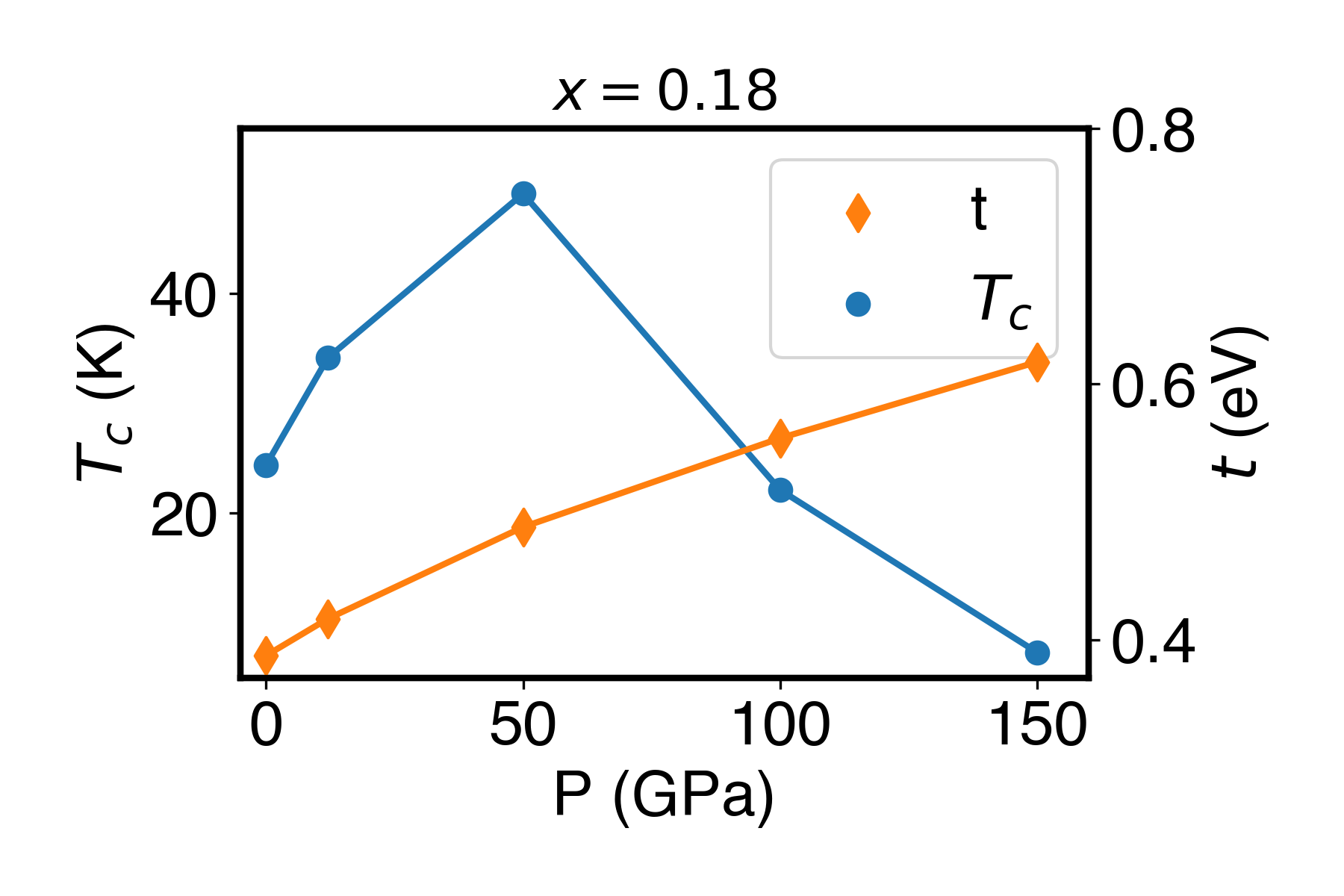}
	\caption{Trend of the superconducting \Tc{} as a function of pressure, compared with the first-nearest-neighbor hopping $t$.}
	\label{fig:tc_and_t_p}
\end{figure}

%\FloatBarrier
%\section{Computational details}
%\label{SSect:computational_details}

%\clearpage

\FloatBarrier
\newpage

\subsection{Tests of the Virtual Crystal Approximation}
\label{subsec:VCAtest}
The Virtual Crystal Approximation should be approached carefully, especially when it is employed to interpolate between two atoms that are not adjacent on the periodic table. 

We performed extensive checks of the quality of the Virtual Crystal Approximation (VCA) to simulate an average mixture of praseodymium (Pr) and strontium (Sr). In the following, we list the tests and their result, and briefly discuss their implication.

\begin{figure}[htb]
	\centering
	\includegraphics[width=0.55\columnwidth]{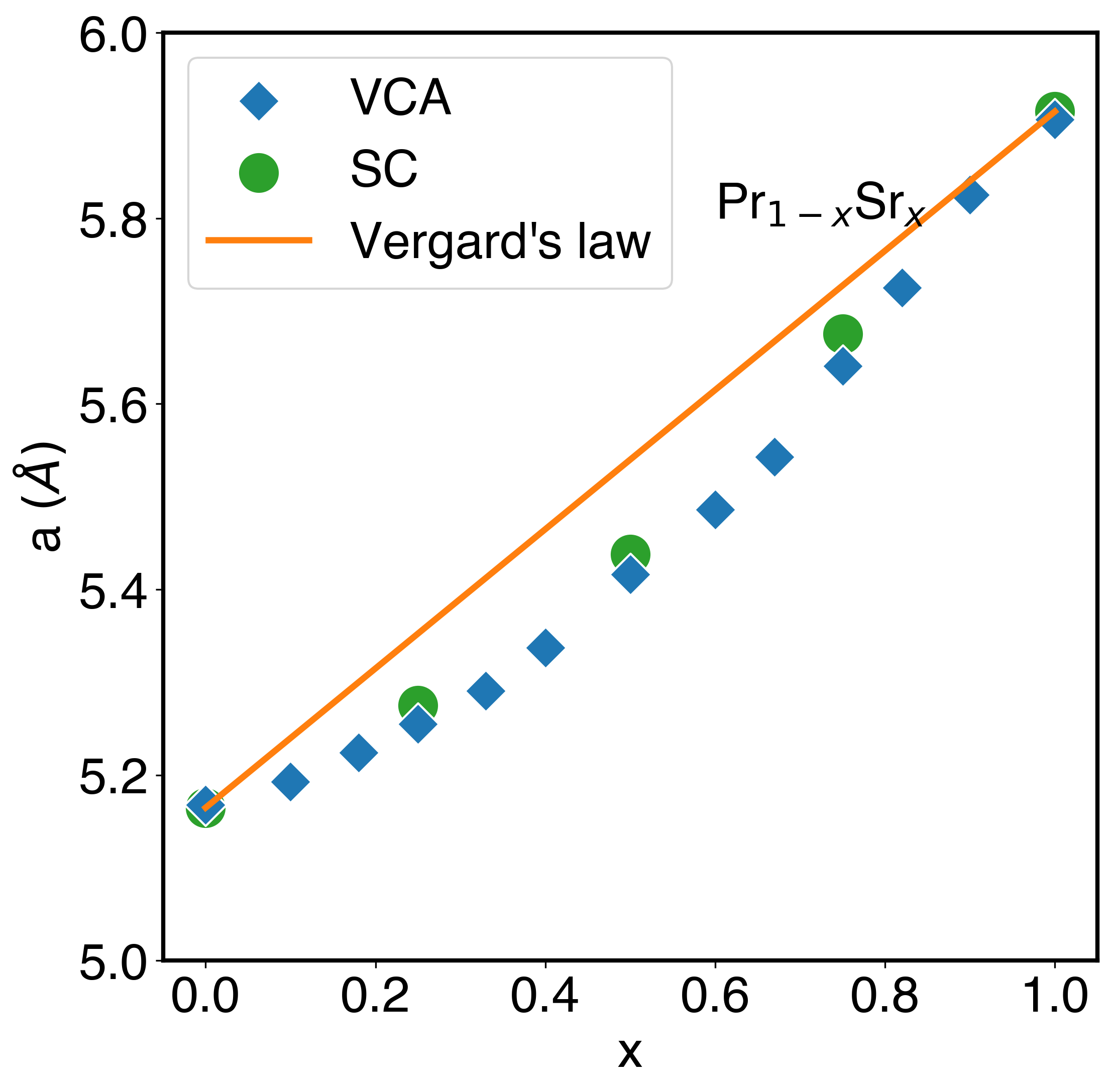}
	\caption{Vergard's Law for a Pr-Sr alloy in the face-centered cubic phase. Blue diamonds, green circles, and the orange line indicate the result obtained using the VCA, a 2$\times$2$\times$2 supercell, and the ideal law, respectively.}
	\label{fig:vergards}
\end{figure}

\paragraph{Vergard's Law}
The first simple test is to check that Vergard's Law is respected \cite{Ashcroft_PRA_2004_Vergard}, i.e. that the lattice parameter of the intermediate alloy is a weighted mean of the two isolate compounds. In Fig. \ref{fig:vergards} we report a comparison between the lattice parameter of a Pr-Sr mixture in a face-centered cubic structure at different concentrations, obtained with the VCA and in a 2$\times$2$\times$2 supercell. We note that not only the law is respected, but even the small deviation from the linear behavior are  matched by the calculations in the supercell, i.e. they are a physical deviation, rather than an artifact of the VCA. 

\paragraph{$c$ axis of Pr$_{0.75}$Sr$_{0.25}$NiO$_2$}
We checked the consistency of the VCA in the specific environment of interest, i.e. the Pr$_{1-x}$Sr$_{x}$NiO$_2$ nickelate. 
A doping of $x = 0.25$ can be simulated in a relatively small 2$\times$2$\times$2 supercell, and its effect on the structural properties compared with the VCA result. 

In Fig. \ref{fig:prsrnio2_vca_test} we show the change in enthalpy of Pr$_{0.75}$Sr$_{0.25}$NiO$_2$ as a function of the $c$ axis using the VCA and in five different supercell configurations. The equilibrium $c$ value thus obtained is identical for all the cases considered, and close to the experimental value \cite{Osada_NanoLett_2020_PrNiO2}, and deviations are only seen away from equilibrium.

\begin{figure}[htb]
	\centering
	\includegraphics[width=0.55\columnwidth]{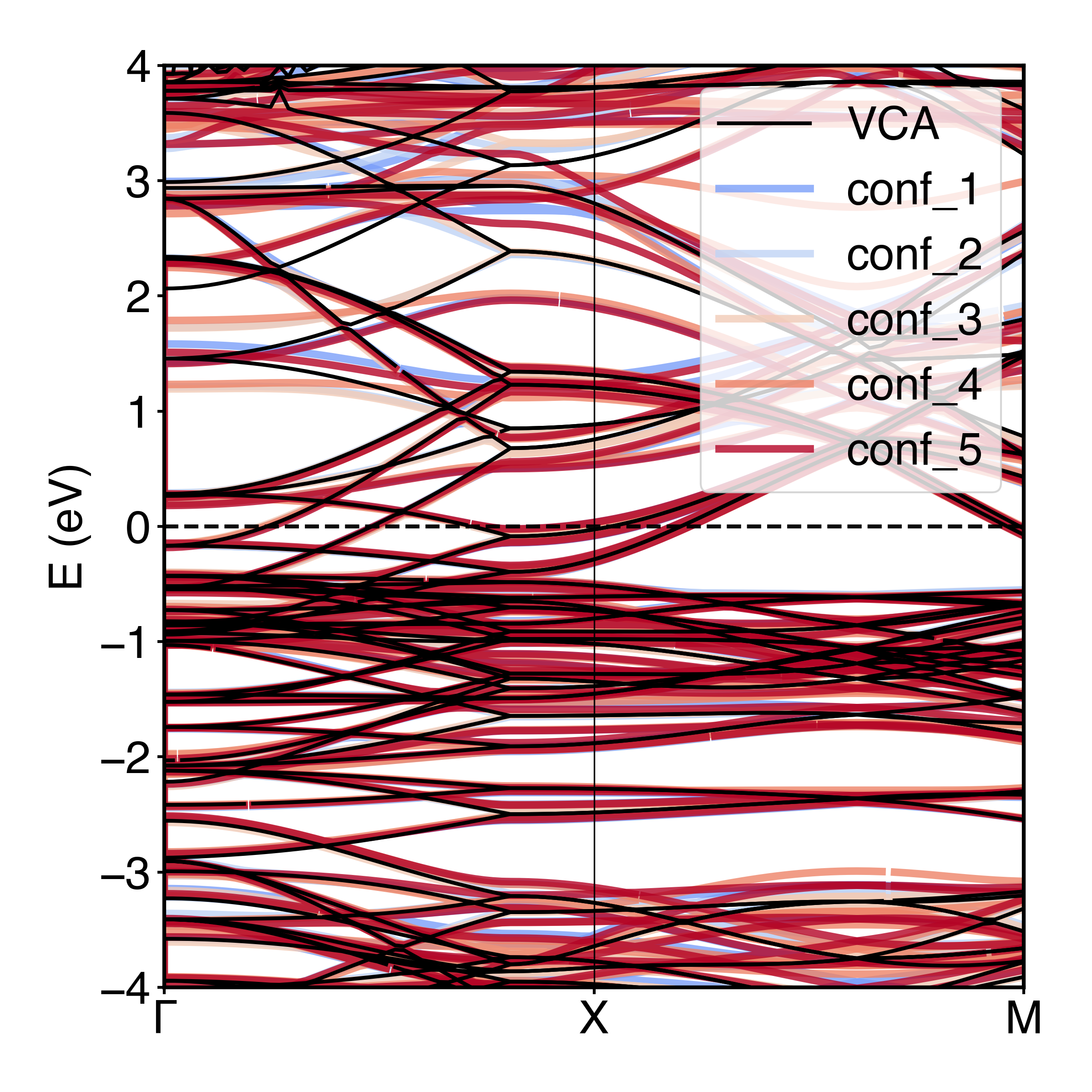}
	\caption{Electronic band structure for Pr$_{0.75}$Sr$_{0.25}$NiO$_2$ using the VCA and in five different supercells. The VCA value is shown as a black line, and the results for the supercells are shown as colored lines.}
	\label{fig:prsrnio2_bands_vca}
\end{figure}

\paragraph{Electronic structure}
The last test involves a direct comparison of the electronic band structure between the VCA case and the five inequivalent supercells at $x = 0.25$. We employed the relaxed structures describe in the previous paragraph, but we note that their lattice parameters are identical within DFT accuracy. In Fig. \ref{fig:prsrnio2_bands_vca} we show the electronic band structure for the VCA and the five different supercells. Also in this case the two results are in excellent agreement, with only a few minute differences due to the symmetry breaking induced by Sr in the supercell, which splits a few bands. 

\FloatBarrier
%\bibliographystyle{apsrev4-2}
%\bibliography{library}
%apsrev4-2.bst 2019-01-14 (MD) hand-edited version of apsrev4-1.bst
%Control: key (0)
%Control: author (72) initials jnrlst
%Control: editor formatted (1) identically to author
%Control: production of article title (-1) disabled
%Control: page (0) single
%Control: year (1) truncated
%Control: production of eprint (0) enabled
%

\end{document}